\documentclass[english,aps,prx,reprint]{revtex4-2}
\usepackage[T1]{fontenc}
\usepackage[latin9]{inputenc}
\usepackage{bm}
\usepackage{amsmath}
\usepackage{amssymb}
\usepackage{graphicx}
\usepackage{wasysym}

\makeatletter

\providecommand{\tabularnewline}{\\}

\usepackage[all]{xy}
\usepackage{mathtools}
\usepackage{variables}
\usepackage[bookmarks=false]{hyperref}

\makeatother

\usepackage{babel}
\begin{document}
\title{Precise ground state of multi-orbital Mott systems via the variational
discrete action theory}
\author{Zhengqian Cheng and Chris A. Marianetti}
\affiliation{Department of Applied Physics and Applied Mathematics, Columbia University,
New York, NY 10027}
\date{\today}
\begin{abstract}
Determining the ground state of multi-orbital Hubbard models is critical
for understanding strongly correlated electron materials, yet existing
methods struggle to simultaneously reach zero temperature and infinite
system size. The \textit{de facto} standard is to approximate a finite
dimension multi-orbital Hubbard model with a $d=\infty$ version,
which can then be formally solved via the dynamical mean-field theory
(DMFT), though the DMFT solution is limited by the state of unbiased
impurity solvers for zero temperature and multiple orbitals. The recently
developed variational discrete action theory (VDAT) offers a new approach
to solve the $d=\infty$ Hubbard model, with a variational ansatz
that is controlled by an integer $\mathcal{N}$, and monotonically
approaches the exact solution at an increasing computational cost.
Here we propose a decoupled minimization algorithm to implement VDAT
for the multi-orbital Hubbard model in $d=\infty$ and study $\mathcal{N}=2-4$
. At $\mathcal{N}=2$, VDAT rigorously recovers the multi-orbital
Gutzwiller approximation, reproducing known results. At $\mathcal{N}=3$,
VDAT precisely captures the competition between the Hubbard $U$,
Hund $J$, and crystal field $\Delta$ in the two orbital Hubbard
model over all parameter space, with a negligible computational cost.
For sufficiently large $U/t$ and $J/U$, we show that $\Delta$ drives
a first-order transition within the Mott insulating regime. In the
large orbital polarization limit with finite $J/U$, we find that
interactions have a nontrivial effect even for small $U/t$. VDAT
will have far ranging implications for understanding multi-orbital
model Hamiltonians and strongly correlated electron materials.
\end{abstract}
\maketitle

\section{Introduction}

The multi-orbital Hubbard model can be considered as a minimal model
for a wide class of strongly correlated electron materials \cite{Imada19981039,Kotliar2006865}.
Due to the complexity of the multi-orbital Hubbard model, it is far
less studied than the single band Hubbard model\cite{Montorsi1992,Gebhard1997,Essler2005,Leblanc2015041041}.
In finite dimensions, there are only a limited number of studies using
state-of-the-art techniques on the multi-orbital Hubbard model, such
as density matrix renormalization group (DMRG) in one dimension\cite{Kaushal2017155111}
and variational quantum Monte-Carlo in two dimensions\cite{Tocchio2016105602,franco2018075117}.
Alternatively, the overwhelming majority of studies focus on the local
physics by approximating the finite dimensional Hubbard model with
a $d=\infty$ counterpart. The $d=\infty$ Hubbard model represents
the essential local physics of the Mott transition, and can be formally
exactly solved using the dynamical mean-field theory (DMFT)\cite{Georges199613,Kotliar200453,Vollhardt20121}.
In this paper, we restrict our attention purely to the ground state
properties at zero temperature. 

The main idea of DMFT is to map the multi-orbital Hubbard model to
a self-consistently determined multi-orbital Anderson impurity model
(AIM), and the self-consistency requires the determination of the
Green's function of the AIM. Two main paradigms to solve the AIM at
zero temperature are Green's function and wave function based techniques.
For Green's function based methods, the zero temperature formalism
is the most natural choice, but this approach is essentially limited
to perturbation theory\cite{Georges199613,Kotliar200453}. Finite
temperature techniques based on the imaginary time Matsubara formalism
can be executed perturbatively or using numerically exact quantum
Monte-Carlo (QMC) based techniques\cite{Gull2011349}. The hybridization
expansion continuous time QMC (CTQMC) \cite{Werner2006076405,Haule2007155113,Gull2011349}
has been extensively used to study the multi-orbital Hubbard model\cite{Werner2006155107,Werner2007126405,Werner2008166405,Poteryaev2008045115,Werner2009115119,Kita2011195130,Hoshino2015247001,Ryee2021206401}.
However, it is computationally expensive to extrapolate to zero temperature.
Wave function based techniques are advantageous in that they naturally
address zero temperature, though they come with their own set of limitations.
The simplest possibility is to discretize the bath of the AIM and
exactly diagonalize the truncated Hamiltonian\cite{Caffarel19941545},
but this approach cannot easily be improved in practice due to the
exponentially increasing computational cost of increasing the bath
size. Both numerical renormalization group (NRG)\cite{Wilson1975773,Wilson1983583,Bulla2008395}
and density matrix renormalization group (DMRG)\cite{Schollwock2005259,Schollwock201196}
allow one to accurately approximate a continuous bath, though each
approach has limitations. NRG uses energy as the criterion to truncate
the Hamiltonian, resulting in an exponential scaling when applied
to a multi-orbital problem\cite{Peters2011075139}. There has been
some success navigating this issue, and there are several studies
executing DMFT using NRG in multi-orbital problems \cite{Pruschke2005217,Stadler2015136401,Kugler2019115159,Kugler2020016401}.
Alternatively, DMRG uses the entanglement entropy as a criteria to
truncate the Hamiltonian, and it should naturally accommodate the
multi-orbital problem. However, given that DMRG only computes the
ground state, additional techniques are needed to obtain the Green's
function\cite{Kuhner1999335,Jeckelmann2002045114,White2004076401,Daley200404005},
which are not well controlled techniques in general. Despite this
limitation, there have been a limited number of applications executing
DMFT using DMRG in two-orbital problems\cite{Fernandez201813,Nunez-fernandez2018121113,Hallberg2020245138,Boidi2021043213}.
In summary, neither NRG nor DMRG studies of the multi-orbital AIM
cover all of parameter space (e.g. the strong interaction regime).
Therefore, there is not yet a universal technique to efficiently study
the ground state of the multi-orbital Hubbard model over all of parameter
space for $d=\infty$, and this seriously limits our ability to study
strongly correlated electron materials. In this paper, we will demonstrate
the VDAT fills this methodological void.

VDAT directly solves the ground state of the Hubbard model in $d=\infty$
without mapping to the AIM\cite{Cheng2021195138,Cheng2021206402}.
VDAT uses a variational ansatz for the many-body density matrix, known
as the sequential product density matrix (SPD), and the accuracy of
the SPD is controlled by an integer $\discn$. Unlike many variational
ansatz, the SPD is unbiased in the sense that it monotonically approaches
the exact solution for increasing $\discn$. In the context of the
Hubbard model, the SPD recovers most well known variational wavefunctions:
$\discn=1$ recovers Hartree-Fock, $\discn=2$ recovers the Gutzwiller
wave function, and $\discn=3$ recovers the Gutzwiller-Baeriswyl and
Baeriswyl-Gutzwiller wavefunctions. For $d=\infty$, VDAT can exactly
evaluate the SPD using the self-consistent canonical discrete action
(SCDA). The computational cost of VDAT grows with $\discn$, at an
exponential scaling for an exact evaluation and a polynomial scaling
for a numerical evaluation using Monte-Carlo, so rapid convergence
with $\discn$ is important if VDAT is to be a practical alternative
to DMFT. Previous work \cite{Cheng2021206402} on the single orbital
AIM on a ring and the $d=\infty$ single orbital Hubbard model are
already well converged for $\discn=3$ as compared to the numerically
exact solution given by DMRG and DMFT solved within NRG, respectively;
with $\discn=4$ pushing the result even closer to the exact solution.
Given that $\discn=2$ recovers the Gutzwiller approximation, which
is already qualitatively reasonable, the great success of $\discn=3$
is not unexpected. In this paper, we will demonstrate that $\discn=3$
maintains a high fidelity in the multi-orbital problem, with complex
local interactions including the Hubbard $U$, Hund $J$, and crystal
field $\Delta$; remedying the known deficiencies of $\discn=2$.
We explicitly show that differences between $\discn=3$ and $\discn=4$
are very small, and comparison to CTQMC extrapolated to zero temperature
yields excellent agreement. Importantly, $\discn=3$ has a similar
computational cost to $\discn=2$, requiring approximately one second
to solve the two-band Hubbard model on a single processor core, orders
of magnitude faster than DMFT solved using QMC based techniques.

It is useful to precisely contrast VDAT within the SCDA to DMFT. Before
the development of VDAT\cite{Cheng2021195138,Cheng2021206402}, DMFT
was the only formalism to exactly solve the Hubbard model in $d=\infty$,
necessitating the use of Green's functions even if one is only concerned
with the ground state properties. VDAT offers a paradigm shift, allowing
the exact solution of the ground state properties of the $d=\infty$
Hubbard model within the wave function paradigm, providing a massive
computational speedup for a given accuracy. For $\discn=2$, the SCDA
provides an alternative approach to the original proof that the Gutzwiller
wave function is exactly evaluated using the Gutzwiller approximation
in $d=\infty$ \cite{Metzner1987121,Metzner19887382,Metzner1989324,Bunemann19977343},
but the SCDA also exactly evaluates the SPD for $\discn\ge3$. From
another viewpoint, the SCDA can be viewed as the integer time analogue
of DMFT, given that the integer time self-energy is assumed to be
local within the SCDA. All of the aforementioned ideas result from
the same simplifications which occur in infinite dimensions. Just
as DMFT can be used as a robust approximation of local physics in
finite dimensions, the SCDA can be applied in the analogous fashion
for determining ground state properties. Moreover, just as DMFT can
be improved in finite dimensions using cluster dynamical mean-field
theory\cite{Maier20051027}, the dynamical vertex approximation\cite{Rohringer2018025003},
dual fermions\cite{Rubtsov2008033101}, etc., VDAT can use the integer
time analogues of these same ideas. 

The structure of this paper is as follows. In Sec. \ref{sec:Variational-Discrete-Action},
we describe the general VDAT formalism, including the SPD and the
DAT. A new derivation for the evaluation of integer time correlation
functions in the compound space is provided, and the gauge freedom
of the SPD is identified and discussed. In Sec. \ref{sec:The-self-consistent-canonical},
we provide an alternate view of the SCDA in terms of two effective
discrete actions subject to self-consistency constraints. Furthermore,
we introduce a decoupled minimization scheme to efficiently execute
the minimization of the variational parameters within the SCDA. The
computational cost of the SCDA is analyzed, and explicit results are
provided. In Sec. \ref{sec:Results-for-Two}, we provide VDAT results
for the two-orbital Hubbard model in $d=\infty$ for a wide variety
of parameters, and compare with published DMFT results. Finally, we
conclude in Sec. \ref{sec:Conclusion-and-Future}. 

\section{Variational Discrete Action Theory \label{sec:Variational-Discrete-Action}}

\subsection{SPD for lattice models}

We begin by reviewing the SPD\cite{Cheng2021195138,Cheng2021206402}
in the context of a lattice model with local interactions, and we
consider a corresponding Hamiltonian defined in an arbitrary lattice
as
\begin{equation}
\hat{H}=\hat{K}+\hat{H}_{loc}=\boldsymbol{\epsilon}\cdot\hat{\boldsymbol{n}}+\sum_{i}\hat{H}_{loc;i},
\end{equation}
where $\boldsymbol{\epsilon}\cdot\hat{\boldsymbol{n}}\equiv\sum_{\ell\ell'}[\boldsymbol{\epsilon}]_{\ell\ell'}[\hat{\boldsymbol{n}}]_{\ell\ell'}$
and $[\hat{\boldsymbol{n}}]_{\ell\ell'}=\hat{a}_{\ell}^{\dagger}\hat{a}_{\ell'}$
and $\ell=1,\dots,L$ labels a complete, orthonormal, single-particle
basis; and $\hat{H}_{loc;i}$ is the local interaction on lattice
site $i$. The ansatz of VDAT is the SPD, and the G-type SPD can be
motivated by considering the following variational wavefunction 
\begin{equation}
|\Psi\rangle=\hat{\mathcal{P}}\left(\bm{\gamma}_{1},u_{1}\right)...\hat{\mathcal{P}}\left(\bm{\gamma}_{M},u_{M}\right)|\Psi_{0}\rangle
\end{equation}
where
\begin{align}
 & \hat{\mathcal{P}}\left(\bm{\gamma}_{j},u_{j}\right)=\exp\left(\bm{\gamma}_{j}\cdot\hat{\bm{n}}\right)\prod_{i}\hat{P}_{i}(u_{j}),\\
 & \hat{P}_{i}\left(u_{j}\right)=\sum_{\Gamma}u_{j,i\Gamma}\hat{P}_{i\Gamma},\label{eq:int_proj_i}
\end{align}
where $j=1,...,M$, the matrices $\bm{\gamma}_{j}$ are Hermitian,
the index $\Gamma$ enumerates a basis of many-body operators $\{\hat{P}_{i\Gamma}\}$
which are Hermitian and local to site $i$, $u_{j}=\{u_{j,i\Gamma}\}$,
and $|\Psi_{0}\rangle$ is a non-interacting wavefunction. The basis
$\{\hat{P}_{i\Gamma}\}$ should be chosen such that the resulting
vector space covers $\exp(\hat{H}_{loc;i})$ for arbitrary interaction
parameters within $\hat{H}_{loc;i}$ (see Section \ref{subsec:Hamiltonian-for-the}
for the choice of $\{\hat{P}_{i\Gamma}\}$ in the two orbital Hubbard
model). The variational parameters are $\bm{\gamma}_{j}$, $u_{j}$,
and the choice of $|\Psi_{0}\rangle$. The integer $M$ sets the accuracy
of the variational wavefunction, and $M\rightarrow\infty$ is guaranteed
to recover the exact wavefunction. In order to execute the variational
theory, the expectation value $\langle\Psi|\hat{H}|\Psi\rangle$ must
be evaluated, and this is best achieved by abstracting to a more general
density matrix ansatz of which this wave function is a special case.
We can rewrite $|\Psi\rangle\langle\Psi|$ as a special case of the
G-type sequential product density matrix\cite{Cheng2021195138,Cheng2021206402}
as 
\begin{align}
\spd & =\hat{\mathcal{P}}\left(\bm{\gamma}_{1},u_{1}\right)\dots\hat{\mathcal{P}}\left(\bm{\gamma}_{\discn},u_{\discn}\right),
\end{align}
where $\mathcal{N}=2M+1$ in general or $\discn=2M$ if one restricts
to $\bm{\gamma}_{1}=\boldsymbol{0}$ for $M>0$. Notice that when
we rewrite $|\Psi_{0}\rangle\langle\Psi_{0}|$ as $\exp\left(\bm{\gamma}_{M+1}\cdot\hat{\bm{n}}\right)$,
the $\bm{\gamma}_{M+1}$ has divergent matrix elements. Therefore,
it is natural to reparametrize $\bm{\gamma}_{\tau}$ using $\bm{\lambda}_{\tau}=(1+\exp(-\bm{\gamma}_{\tau})^{T})^{-1}$,
and $\bm{\lambda}_{M+1}$ is the single particle density matrix of
$|\Psi_{0}\rangle$. The variational parameters are then $\lambda=\{\boldsymbol{\lambda}_{1},\dots,\boldsymbol{\lambda}_{M+1}\}$
and $u=\{u_{1},\dots,u_{M}\}$, and the remaining parameters are given
as $\bm{\lambda}_{M+1+k}\equiv\bm{\lambda}_{M+1-k}$, $u_{M+k}\equiv u_{M+1-k}$,
where $k>0$, with $u_{0}$ chosen such that $\hat{P}_{i}\left(u_{0}\right)=\hat{1}$.
While we have focused on the G-type SPD, which is used in our present
calculations, it is worth noting that there is a second class of SPD
denoted as B-type \cite{Cheng2021195138}. 

The variational principle dictates that the ground state energy is
evaluated as 
\begin{equation}
E=\min_{\lambda u}\langle\hat{H}\rangle_{\spd(\lambda,u)},
\end{equation}
where $\langle\hat{O}\rangle_{\hat{\rho}}\equiv\textrm{Tr}(\hat{\rho}\hat{O})/\textrm{Tr}(\hat{\rho})$.
The accuracy of the SPD is controlled by $\discn$, and the error
will monotonically decrease with increasing $\discn$. There are two
main challenges posed by the SPD ansatz: exactly evaluating $\langle\hat{H}\rangle_{\spd(\lambda,u)}$
and minimizing over the sets of variational parameters $\lambda$
and $u$. In the case of a $d=\infty$ lattice, we previously proved
that the SCDA can be used to exactly evaluate $\langle\hat{H}\rangle_{\spd(\lambda,u)}$\cite{Cheng2021195138},
and here we demonstrate that the SCDA can be executed efficiently
for the two band Hubbard model. 

It should be emphasized that the SPD is defined by the sequence $(\hat{\mathcal{P}}_{1},\dots,\hat{\mathcal{P}}_{\discn})$,
where $\hat{\mathcal{P}}_{\tau}\equiv\hat{\mathcal{P}}\left(\bm{\gamma}_{\tau},u_{\tau}\right)$,
and therefore for $\discn>1$ there are always distinct SPD's that
correspond to an equivalent many-body density matrix, which we refer
to as gauge equivalent. The SPD gives rise to the notion of integer
time correlation functions of the form $\text{Tr}(\hat{\mathcal{P}}_{1}\hat{O}_{1}\dots\hat{\mathcal{P}}_{\discn}\hat{O}_{\mathcal{N}})/\textrm{Tr}(\spd)$.
While it may not be immediately obvious why this correlation function
is relevant, integer time correlation function naturally emerge when
constructing a diagrammatic expansion and when evaluating the derivatives
of the energy with respect to the variational parameters\cite{Cheng2021195138}. 

\subsection{The discrete action theory represented in the compound space}

The discrete action theory (DAT) is a general formalism to evaluate
integer time correlation functions of the SPD \cite{Cheng2021195138}.
It is convenient to represent the DAT in a compound space $\mathbb{H}_{c}=\otimes_{\tau=1}^{\discn}\mathbb{H}$,
where $\mathbb{H}$ is the original Fock space. Each pair of operators
$\hat{a}{}_{\ell}^{\dagger}$ and $\hat{a}{}_{\ell}$ can be promoted
into $\mathbb{H}_{c}$ as $\discn$ distinct pairs of operators with
integer time index $\tau=1,\dots,\discn$, denoted $\barhat[a]_{\ell}^{\dagger(\tau)}$
and $\barhat[a]_{\ell}^{(\tau)}$, which are defined by the canonical
anti-commutation relations $\{\barhat[a]_{\ell}^{\dagger(\tau)},\barhat[a]_{\ell'}^{(\tau')}\}=\delta_{\ell\ell'}\delta_{\tau\tau'}$
and $\{\barhat[a]_{\ell}^{(\tau)},\barhat[a]_{\ell'}^{(\tau')}\}=0$.
Any operator $\hat{O}=f(\{\hat{a}{}_{\ell}^{\dagger}\},\{\hat{a}{}_{\ell}\})$
can be promoted to $\mathbb{H}_{c}$ with time index $\tau$ as $\barhat[O]^{\left(\tau\right)}=f(\{\barhat[a]{}_{\ell}^{\dagger(\tau)}\},\{\barhat[a]_{\ell}^{(\tau)}\})$,
and this implies that promotion preserves the algebraic structure
of the operator. The utility of the compound space and promoted operators
can be seen from the following identity, 
\begin{equation}
\frac{\text{Tr}(\hat{\mathcal{P}}_{1}\hat{O}_{1}\dots\hat{\mathcal{P}}_{\discn}\hat{O}_{\mathcal{N}})}{\text{Tr}(\hat{\mathcal{P}}_{1}\dots\hat{\mathcal{P}}_{\discn})}=\langle\barhat[O]_{1}^{\left(1\right)}\dots\barhat[O]_{\discn}^{\left(\discn\right)}\rangle_{\spdcs},\label{eq:equivalence_correlation_function}
\end{equation}
where the left hand side of the equation is a general integer time
correlation function and the right hand side is the corresponding
observable evaluated under the many-body operator $\spdcs$ in $\mathbb{H}_{c}$,
and we refer to $\spdcs$ as the discrete action. The discrete action
is defined as $\spdcs\equiv\barhat[Q]\prod_{\tau}\barhat[\mathcal{P}]_{\tau}^{(\tau)}$,
where $\barhat[Q]$ is a unitary operator in $\mathbb{H}_{c}$ defined
as 

\begin{equation}
\barhat[Q]\equiv\exp\left(\sum_{\ell\left(\tau\neq\tau'\right)}\frac{\pi}{\discn\sin\left(\pi\left(\tau-\tau'\right)/\discn\right)}\barhat[a]_{\ell}^{\dagger\left(\tau\right)}\barhat[a]_{\ell}^{\left(\tau'\right)}\right).
\end{equation}
We can show that $\barhat[Q]^{-1}\barhat[a]_{\ell}^{\dagger(\tau)}\barhat[Q]=-\barhat[a]_{\ell}^{\dagger(\tau+1)}$
and $\barhat[Q]^{-1}\barhat[a]_{\ell}^{(\tau)}\barhat[Q]=-\barhat[a]_{\ell}^{(\tau+1)}$,
where $\barhat[a]_{\ell}^{\dagger(\discn+1)}\equiv-\barhat[a]_{\ell}^{\dagger(1)}$,
and thus we refer to $\barhat[Q]$ as the integer time translation
operator, which encodes the intrinsic time correlation of the SPD
and only depends on $\mathcal{N}$ (see Section \ref{sec:Appendix}
for derivation).

The equivalence in Eq. \ref{eq:equivalence_correlation_function}
was first proved using an explicit matrix representation\cite{Cheng2021195138},
and here we provide an alternative proof based on a diagrammatic expansion.
We first prove that for arbitrary $\hat{O}_{\tau}$, 
\begin{equation}
\langle\hat{O}_{1}\dots\hat{O}_{\discn}\rangle_{\hat{1}}=\langle\barhat[O]_{1}^{\left(1\right)}\dots\barhat[O]_{\discn}^{(\discn)}\rangle_{\barhat[Q]}.\label{eq:equivalence_relation2}
\end{equation}
Notice that $\hat{1}$ and $\barhat[Q]$ are non-interacting operators
in the original space and compound space, respectively, so Wick's
theorem may be employed in both cases. Given that the promotion does
not change the algebraic structure, both expectation values will yield
the same diagrams with corresponding contractions. Therefore, the
only point to be verified is that all contractions that appear in
the diagrammatic expansion are equivalent, which will be satisfied
if
\begin{equation}
\langle\barhat[A]_{\ell;\eta}^{\left(\tau\right)}\barhat[A]_{\ell';\eta'}^{\left(\tau'\right)}\rangle_{\barhat[Q]}=\langle\hat{A}_{\ell;\eta}\hat{A}_{\ell';\eta'}\rangle_{\hat{1}}
\end{equation}
for all $\tau\leq\tau'$, where $\hat{A}_{\ell;0}\equiv\hat{a}_{\ell}^{\dagger}$
and $\hat{A}_{\ell;1}\equiv\hat{a}_{\ell}.$ First, one can directly
compute $\langle\hat{A}_{\ell;\eta}\hat{A}_{\ell';\eta'}\rangle_{\hat{1}}=\frac{1}{2}\delta_{\ell\ell'}\delta_{\left|\eta-\eta'\right|,1}$.
Second, the definition of $\barhat[Q]$ gives $\langle\barhat[a]_{\ell}^{\left(\tau\right)\dagger}\barhat[a]_{\ell'}^{\left(\tau'\right)}\rangle_{\barhat[Q]}=\frac{1}{2}\text{sign\ensuremath{\left(\tau'-\tau+\frac{1}{2}\right)}}\delta_{\ell\ell'}$
and $\langle\barhat[a]_{\ell}^{\left(\tau\right)}\barhat[a]_{\ell'}^{\left(\tau'\right)}\rangle_{\barhat[Q]}=0$,
which proves Eq. \ref{eq:equivalence_relation2}. Equation \ref{eq:equivalence_relation2}
can now be applied in two instances using $\hat{O}_{\tau}\rightarrow\hat{\mathcal{P}}_{\tau}$
and $\hat{O}_{\tau}\rightarrow\mathcal{\hat{P}}_{\tau}\hat{O}_{\tau}$,
respectively, and subsequently dividing the latter by the former,
which will yield Eq. \ref{eq:equivalence_correlation_function} given
that $\barhat[\mathcal{P}]_{\tau}^{\left(\tau\right)}$ are bosonic
and commute with any operator in a different integer time.

Using Eq. \ref{eq:equivalence_correlation_function}, the ground state
energy under $\spd$ can be equivalently evaluated in the compound
space $\mathbb{H}_{c}$ as 
\begin{equation}
\langle\hat{H}\rangle_{\spd(\lambda,u)}=\langle\barhat[H]^{\left(\mathcal{N}\right)}\rangle_{\spdcs\left(\lambda,u\right)},\label{eq:energy_compond}
\end{equation}
where the discrete action can be rearranged into a product of a noninteracting
and interacting part\cite{Cheng2021195138}, given as 

\begin{equation}
\spdcs\left(\lambda,u\right)=\spdcs_{0}\left(\lambda\right)\prod_{i\tau}\barhat[P]_{i}^{\left(\tau\right)}\left(u_{\tau}\right),\label{eq:spdcs}
\end{equation}
where

\begin{equation}
\spdcs_{0}\left(\lambda\right)=\barhat[Q]\exp\left(-\sum_{\tau}\ln\left(\bm{\lambda}_{\tau}^{-1}-\boldsymbol{1}\right)^{T}\cdot\barhat[\bm{n}]^{(\tau)}\right).
\end{equation}
It should be emphasized that the partitioning in Eq. \ref{eq:spdcs}
is only possible because projectors from different integer time steps
commute with each other.

A direct evaluation of Eq. \ref{eq:energy_compond} in the compound
space has no obvious advantage, but the methodological advantage is
that $\spdcs_{0}$ can be used as a non-interacting starting point
and $\barhat[P]$ can be treated as a perturbation. Therefore, we
can generalize the usual Green's function techniques to the integer
time case\cite{Cheng2021195138}, and an important example is the
discrete Dyson equation 
\begin{equation}
\boldsymbol{g}^{-1}-\boldsymbol{1}=\left(\boldsymbol{g}_{0}^{-1}-\boldsymbol{1}\right)\boldsymbol{S},\label{eq:discrete_dyson}
\end{equation}
where $\text{\ensuremath{\bm{g}}}=\left\langle \barhat[\bm{n}]\right\rangle _{\spdcs}$,
$\text{\ensuremath{\bm{g}}}_{0}=\left\langle \barhat[\bm{n}]\right\rangle _{\spdcs_{0}}$,
$[\barhat[\bm{n}]]_{\ell\tau,\ell'\tau'}=\barhat[a]_{\ell}^{\dagger(\tau)}\barhat[a]_{\ell'}^{(\tau')},$
and $\boldsymbol{S}$ is the exponential form of the integer time
self-energy. The discrete Dyson equation, which is a matrix equation
of dimension of $L\discn\times L\discn$, exactly relates the interacting
and noninteracting integer time Green's function via $\boldsymbol{S}$.
It should be emphasized that the discrete Dyson equation is not equivalent
to discretizing the usual Dyson equation (see Sections IVB and IVC
in Ref. \cite{Cheng2021195138}). 

\subsection{Gauge freedom of SPD and VDAT}

Recall that the SPD is defined by the sequence $(\hat{\mathcal{P}}_{1},\dots,\hat{\mathcal{P}}_{\discn})$,
and therefore for $\discn>1$ there are always distinct SPD's that
correspond to an equivalent many-body density matrix, which we refer
to as gauge equivalent. A gauge transformation can be defined by the
transformation $\hat{\mathcal{P}}_{\tau}\rightarrow\hat{\mathcal{P}}_{\tau}'$
such that $\hat{\mathcal{P}}_{1}\dots\mathcal{\hat{P}}_{\discn}=\hat{\mathcal{P}}'_{1}\dots\mathcal{\hat{P}}'_{\discn}$.
Therefore, this gauge freedom must be fixed in order to avoid numerical
instabilities. To illustrate the gauge freedom, we consider an $\mathcal{N}=2$,
G-type SPD $\spd=\hat{P}_{1}\hat{K}_{2}\hat{P}_{1}^{\dagger}$ and
a gauge transformation $\hat{P}'_{1}=\hat{P}_{1}\hat{N}$ and $\hat{K}'_{2}=\hat{N}^{-1}\hat{K}_{2}(\hat{N}^{\dagger})^{-1}$,
resulting in $\spd'=\hat{P}'_{1}\hat{K}'_{2}(\hat{P}'_{1})^{\dagger}$;
where $\hat{N}$ is a general non-interacting operator. The kinetic
projector and local projector are transformed into new forms, which
yields different integer time Green's functions and self-energies,
but will yield the same static expectation values (i.e. where all
observables are measured in the last integer time step). To consider
how $\boldsymbol{g}$ and $\boldsymbol{S}$ change, consider the the
explicit example of $\hat{N}=\hat{N}^{\dagger}=\exp(\mu\sum_{\ell}\hat{n}_{\ell})$,
where $\boldsymbol{g}$ is changed as
\begin{equation}
\bm{g}'=\textrm{diag}(\exp\left(\mu\boldsymbol{1}\right),\boldsymbol{1})\hspace{0.7mm}\bm{g}\hspace{0.7mm}\textrm{diag}(\exp\left(-\mu\boldsymbol{1}\right),\boldsymbol{1}),
\end{equation}
and $\boldsymbol{S}$ is changed as
\begin{equation}
\bm{S}'=\textrm{diag}(\boldsymbol{1},\exp\left(-\mu\boldsymbol{1}\right))\hspace{0.7mm}\bm{S}\hspace{0.7mm}\textrm{diag}(\exp\left(-\mu\boldsymbol{1}\right),\boldsymbol{1}).
\end{equation}
One possible way to constrain this gauge is by requiring $\left|\bm{S}\right|=1$.
For the case of $\discn=3$ G-type SPD's, the gauge transformation
has the form $\hat{K}_{1}\hat{P}_{1}\hat{K}_{2}\hat{P}_{1}^{\dagger}\hat{K}_{1}=\hat{K}'_{1}\hat{P}'_{1}\hat{K}'_{2}(\hat{P}'_{1})^{\dagger}\hat{K}'_{1}$,
yielding two possibilities. First, we have $\hat{K}'_{1}=\hat{K}_{1}$
and $\hat{P}_{1}'=\hat{P}_{1}\hat{N}$ and $\hat{K}'_{2}=\hat{N}^{-1}\hat{K}_{2}\left(\hat{N}^{\dagger}\right)^{-1}$,
yielding 
\begin{align}
 & \boldsymbol{g}'=\textrm{diag}(\exp(\mu\boldsymbol{1}),\boldsymbol{1},\boldsymbol{1})\boldsymbol{\bm{g}}\textrm{diag}(\exp(-\mu\boldsymbol{1}),\boldsymbol{1},\boldsymbol{1}),\\
 & \bm{S}'=\textrm{diag}(\boldsymbol{1},\exp(-\mu\boldsymbol{1}),\boldsymbol{1})\boldsymbol{S}\textrm{diag}(\exp(-\mu\boldsymbol{1}),\boldsymbol{1},\boldsymbol{1}).
\end{align}
Second, we have $\hat{K}'_{1}=\hat{K}_{1}\hat{N}^{-1}$ and $\hat{P}_{1}'=\hat{N}\hat{P}_{1}$
and $\hat{K}'_{2}=\hat{K}_{2}$, yielding
\begin{align}
 & \boldsymbol{g}'=\textrm{diag}(\boldsymbol{1},\exp(\mu\boldsymbol{1}),\boldsymbol{1})\boldsymbol{\bm{g}}\textrm{diag}(\boldsymbol{1},\exp(-\mu\boldsymbol{1}),\boldsymbol{1}),\\
 & \bm{S}'=\textrm{diag}(\exp(-\mu\boldsymbol{1}),\boldsymbol{1},\boldsymbol{1})\boldsymbol{S}\textrm{diag}(\boldsymbol{1},\exp(-\mu\boldsymbol{1}),\boldsymbol{1}).
\end{align}
Both gauges may be constrained by requiring $\left|\bm{S}\right|=1$
and requiring the determinants of the $(1,2)$ and $(2,1)$ integer
time sub-blocks of $\boldsymbol{S}$ are the negative of each other.
The case of $\discn=4$ is discussed in Supplemental Material\cite{supplementary}.

\section{The self-consistent canonical discrete action theory\label{sec:The-self-consistent-canonical}}

\subsection{General formulation of the SCDA}

The key idea of the SCDA\cite{Cheng2021195138,Cheng2021206402} is
to approximately compute the total energy with two effective discrete
actions that are determined self-consistently, and a key feature of
the SCDA is that it becomes exact for $d=\infty$\cite{Cheng2021195138}.
The kinetic energy is determined by $\barhat[\rho]_{K}$, which approximates
the exact interacting projector by a non-interacting operator parameterized
by $S=\{\bm{S}_{i}\}$, where $\bm{S}_{i}$ is local to site $i$
and has dimension $N_{i}\discn\times N_{i}\discn$ where $N_{i}$
is the number of spin orbitals associated with site $i$. Alternatively,
the local interaction energy is determined by $\barhat[\rho]_{loc}$,
which approximates the non-interacting discrete action and is parametrized
by $\mathcal{G}=\{\bm{\mathcal{G}}_{i}\}$, where $\bm{\mathcal{G}}_{i}$
is local to site $i$ and has dimension $N_{i}\discn\times N_{i}\discn$.
Finally, $S$ and $\mathcal{G}$ are uniquely determined by the variational
parameters $\lambda$ and $u$ through the self-consistency of the
local integer time Green's function and the discrete Dyson equation.
Mathematically, this procedure is described by 
\begin{equation}
E\left(\lambda,u\right)=\langle\barhat[K]^{\left(\mathcal{N}\right)}\rangle_{\barhat[\rho]_{K}}+\langle\barhat[H]_{loc}^{\left(\mathcal{N}\right)}\rangle_{\barhat[\rho]_{loc}},
\end{equation}
where 

\begin{align}
\barhat[\rho]_{K} & =\spdcs_{0}\left(\lambda\right)\prod_{i}\exp(-\ln\bm{S}_{i}^{T}\cdot\barhat[\bm{n}]_{i}),\label{eq:rhok}\\
\barhat[\rho]_{loc} & =\prod_{i}\big(\barhat[\rho]_{loc;i}^{0}(\bm{\mathcal{G}}_{i})\prod_{\tau}\hat{P}_{i}^{(\tau)}(u_{\tau})\big),\label{eq:rholoc}
\end{align}
where $[\barhat[\bm{n}]_{i}]_{m\tau,m'\tau'}=\barhat[a]_{im}^{\dagger(\tau)}\barhat[a]_{im'}^{(\tau')}$
and $m$ is an index which labels a spin orbital associated with site
$i$, and $\barhat[\rho]_{loc,i}^{0}(\bm{\mathcal{G}}_{i})=\exp(-\ln\left(\bm{\mathcal{G}}{}_{i}^{-1}-\boldsymbol{1}\right)^{T}\cdot\barhat[\bm{n}]{}_{i})$.
Finally, $S$ and $\mathcal{G}$ can be determined by the following
two conditions for all $i$:

\begin{align}
\left(\boldsymbol{g}_{i}^{-1}-\boldsymbol{1}\right)=\left(\bm{\mathcal{G}}_{i}^{-1}-\boldsymbol{1}\right)\boldsymbol{S}_{i},\hspace{2em}\bm{g}_{i}=\bm{g}'_{i},\label{eq:dysonlocal}
\end{align}
where $\boldsymbol{g}_{i}=\left\langle \barhat[\boldsymbol{n}]_{i}\right\rangle _{\barhat[\rho]_{loc}}$
and $\boldsymbol{g}'_{i}=\left\langle \barhat[\boldsymbol{n}]_{i}\right\rangle _{\barhat[\rho]_{K}}$.

The preceding discussion fully defines the SCDA algorithm, and now
we consider how to evaluate the expectation of an arbitrary operator
under $\barhat[\rho]_{K}$ or $\barhat[\rho]_{loc}$. Given that $\barhat[\rho]_{K}$
is noninteracting, it is straightforward to evaluate the expectation
value. For $\barhat[\rho]_{loc}$, we first evaluate a local operator
$\barhat[O]_{i}$ as
\begin{align}
\langle\barhat[O]_{i}\rangle_{\barhat[\rho]_{loc}} & =\frac{\sum_{\{\Gamma_{\tau}\}}\langle(\prod_{\tau}\barhat[P]_{i\Gamma_{\tau}}^{\left(\tau\right)})\barhat[O]_{i}\rangle_{\barhat[\rho]_{loc,i}^{0}}\prod_{\tau}u_{\tau,i\Gamma_{\tau}}}{\sum_{\{\Gamma_{\tau}\}}\langle\prod_{\tau}\barhat[P]_{i\Gamma_{\tau}}^{\left(\tau\right)}\rangle_{\barhat[\rho]_{loc,i}^{0}}\prod_{\tau}u_{\tau,i\Gamma_{\tau}}},\label{eq:local_operator_expansion_u}
\end{align}
where $\langle(\prod_{\tau}\barhat[P]_{i\Gamma_{\tau}}^{\left(\tau\right)})\barhat[O]_{i}\rangle_{\barhat[\rho]_{loc,i}^{0}}$
and $\langle\prod_{\tau}\barhat[P]_{i\Gamma_{\tau}}^{\left(\tau\right)}\rangle_{\barhat[\rho]_{loc,i}^{0}}$
can be evaluated using Wick's theorem, resulting in a finite polynomial
in terms of the entries of $\boldsymbol{\mathcal{\bm{G}}}{}_{i}$.
For a product of local operators on distinct sites, we have $\langle\barhat[A]\dots\barhat[B]\rangle_{\barhat[\rho]_{loc}}=\langle\barhat[A]\rangle_{\barhat[\rho]_{loc}}\dots\langle\barhat[B]\rangle_{\barhat[\rho]_{loc}}$.
Finally, an arbitrary operator can be written as a sum over products
of local operators, allowing the evaluation of any observable. However,
given that $S$ and $\mathcal{G}$ are implicit functions of $\lambda$
and $u$, taking the gradient of the energy with respect to $\lambda$
and $u$ is nontrivial. This issue will be circumvented using the
decoupled minimization scheme presented below.

\subsection{Decoupled minimization algorithm for the SCDA\label{subsec:Decoupled-minimization-algorithm}}

Given that $\bm{S}{}_{i}$ and $\bm{\mathcal{G}}_{i}$ are constrained
by the discrete Dyson equation, one of them can be eliminated. One
can begin with either $\bm{\mathcal{G}}_{i}$ or $\bm{S}{}_{i}$,
and this will yield distinct but equivalent decoupled minimization
algorithms. Here we start with $\bm{\mathcal{G}}_{i}$, and then $\barhat[\rho]_{loc}\left(u,\mathcal{G}\right)$
can be determined from Eq\ref{eq:rholoc}, which determines $\boldsymbol{g}_{i}\left(u,\mathcal{G}\right)=\left\langle \barhat[\boldsymbol{n}]_{i}\right\rangle _{\barhat[\rho]_{loc}}$and
$E_{loc}\left(u,\mathcal{G}\right)=\langle\barhat[H]_{loc}^{\left(\mathcal{N}\right)}\rangle_{\barhat[\rho]_{loc}}$.
Using the discrete Dyson equation, we have 
\begin{equation}
\bm{S}{}_{i}\left(u,\mathcal{G}\right)=\left(\bm{\mathcal{G}}_{i}^{-1}-\bm{1}\right)^{-1}\left(\bm{g}_{i}{}^{-1}-\bm{1}\right).
\end{equation}
The $\barhat[\rho]_{K}\left(\lambda,u,\mathcal{G}\right)$ can be
determined from Eq \ref{eq:rhok}, which determines $\boldsymbol{g}'_{i}\left(\lambda,u,\mathcal{G}\right)=\left\langle \barhat[\boldsymbol{n}]_{i}\right\rangle _{\barhat[\rho]_{K}}$
and $K\left(\lambda,u,\mathcal{G}\right)=\langle\barhat[K]^{\left(\mathcal{N}\right)}\rangle_{\barhat[\rho]_{K}}$.
Finally, we compute the total energy as $E\left(\lambda,u,\mathcal{G}\right)=K+E_{loc}$,
and the constraint function is $\bm{\Delta}_{i}\left(\lambda,u,\mathcal{G}\right)\equiv\bm{g}_{i}-\bm{g}'_{i}$.
In summary, the problem can be cast as 
\begin{equation}
E=\begin{array}{rl}
\min & E\left(\lambda,u,\mathcal{G}\right)\\[0.5em]
\textrm{subject to} & \bm{\Delta}_{i}\left(\lambda,u,\mathcal{G}\right)=\boldsymbol{0},\hspace{1em}i=1,\dots,N_{\textrm{site}}
\end{array}
\end{equation}
The constraint can be implemented by assuming that $\bm{g}'_{i}$
and $\bm{S}{}_{i}$ are constant, allowing for a solution for $\bm{\mathcal{G}}_{i}$
as
\begin{equation}
\bm{\mathcal{G}}'{}_{i}=(\boldsymbol{1}+\left(\bm{g}_{i}'{}^{-1}-1\right)\bm{S}_{i}^{-1})^{-1}.\label{eq:update_g}
\end{equation}
The new $\bm{\mathcal{G}}'{}_{i}$ can then be used to start a new
iteration, and this process will be iterated until self-consistency
is achieved. 

To minimize the variational parameters $\lambda$ and $u$, we begin
by computing the first derivative of $E\left(\lambda,u,\mathcal{G}\right)$
with respect to $\lambda$ for a fixed $\Delta=\{\boldsymbol{\Delta}_{i}\}$
and $u$, given as 

\begin{align}
\frac{dE}{d\lambda} & =\frac{\partial K}{\partial\lambda}+\sum_{i}\frac{\partial E}{\partial\bm{\mathcal{G}}_{i}}\cdot\frac{\partial\bm{\mathcal{G}}_{i}}{\partial\lambda}\big|_{\Delta},
\end{align}
where
\begin{equation}
\frac{\partial E}{\partial\bm{\mathcal{G}}_{i}}\cdot\frac{\partial\bm{\mathcal{G}}_{i}}{\partial\lambda}\big|_{\Delta}\equiv\sum_{mm'}\frac{\partial E}{\partial[\bm{\mathcal{G}}_{i}]_{mm'}}\frac{\partial[\bm{\mathcal{G}}_{i}]_{mm'}}{\partial\lambda}\big|_{\Delta}.
\end{equation}
The above notation indicates how contraction is performed between
the respective tensors. Using $\frac{\partial\bm{\mathcal{G}}_{i}}{\partial\lambda}\big|_{\Delta}=-\sum_{i'}\frac{\partial\bm{\mathcal{G}}_{i}}{\partial\bm{\Delta}_{i'}}\cdot\frac{\partial\bm{\Delta}_{i'}}{\partial\bm{g}'}\cdot\frac{\partial\bm{g}'}{\partial\lambda}$
and $\partial K/\partial\lambda=\frac{\partial}{\partial\lambda}\left(\boldsymbol{\epsilon}\cdot\bm{n}\right)$,
where for a given $i$ and $i'$ the derivative identity $\frac{\partial[\bm{\mathcal{G}}_{i}]_{m_{1}m_{2}}}{\partial\bm{\Delta}_{i'}}\cdot\frac{\partial\bm{\Delta}_{i'}}{\partial[\bm{\mathcal{G}}_{i}]_{m_{1}'m_{2}'}}=\delta_{m_{1}m_{1}'}\delta_{m_{2}m_{2}'}$
can be used to obtain $\partial\bm{\mathcal{G}}_{i}/\partial\bm{\Delta}_{i'}$
from $\partial\bm{\Delta}_{i'}/\partial\bm{\mathcal{G}}_{i}$, an
effective potential $\bm{v}_{K,i}=\sum_{i'}(\partial E/\partial\bm{\mathcal{G}}_{i'})\cdot(\partial\bm{\mathcal{G}}_{i'}/\partial\bm{\Delta}_{i})$
can be constructed in the compound space such that
\begin{equation}
\frac{dE}{d\lambda}=\frac{\partial}{\partial\lambda}\langle\boldsymbol{\epsilon}\cdot\barhat[\bm{n}]^{\left(\discn\right)}+\sum_{i}\text{\ensuremath{\bm{v}}}_{K,i}\cdot\barhat[\bm{n}]_{i}\rangle_{\barhat[\rho]_{K}\left(\lambda,S\right)},
\end{equation}
where $S$ and $\boldsymbol{v}_{K,i}$ are held constant when taking
the derivative. This allows $\lambda$ to be updated as 
\begin{equation}
\lambda'=\text{argmin}_{\lambda}\langle\boldsymbol{\epsilon}\cdot\barhat[\bm{n}]^{\left(\discn\right)}+\sum_{i}\text{\ensuremath{\bm{v}}}_{K,i}\cdot\barhat[\bm{n}]_{i}\rangle_{\barhat[\rho]_{K}\left(\lambda,S\right)},\label{eq:update_lambda}
\end{equation}
where $S$ and $\boldsymbol{v}_{K,i}$ are held constant when minimizing
over $\lambda$. To compute $\bm{v}_{K}$, we use the automatic differentiation
technique in the forward mode. An analogous procedure can be used
to update $u$ as

\begin{align}
 & u'=\text{argmin}_{u}\langle\barhat[H]_{loc}^{\left(\discn\right)}+\sum_{i}\bm{v}_{loc,i}\cdot\barhat[\bm{n}]_{i}\rangle_{\barhat[\rho]_{loc}\left(u,\mathcal{G}\right)},\label{eq:update_u}\\
 & \bm{v}_{loc,i}=\frac{\partial K}{\partial\bm{g}_{i}}-\sum_{i'i''}\frac{\partial E}{\partial\bm{\mathcal{G}}_{i'}}\cdot\frac{\partial\bm{\mathcal{G}}_{i'}}{\partial\bm{\Delta}_{i''}}\cdot\frac{\partial\bm{\Delta}_{i''}}{\partial\bm{g}_{i}},
\end{align}
where $\mathcal{G}$ and $\bm{v}_{loc,i}$ are held constant when
minimizing over $u$. Finally, $\mathcal{G}$, $\lambda$, and $u$
can be updated using Eqns. \ref{eq:update_g}, \ref{eq:update_lambda},
and \ref{eq:update_u} in each iteration, and when all quantities
converge, the constraint has been satisfied while minimizing over
all variational parameters. 

\subsubsection{Translation Symmetry\label{subsec:Translation-Symmetry}}

In many cases, we will be solving a Hamiltonian that is invariant
to translation symmetry, which will dramatically reduce the computational
cost within the SCDA. Translation symmetry dictates that $\boldsymbol{\mathcal{G}}_{i}$
and $u_{\tau,i\Gamma}$ are independent of $i$, and therefore we
make the simplification $\boldsymbol{\mathcal{G}}_{i}\rightarrow\boldsymbol{\mathcal{G}}$
and $u_{\tau,i\Gamma}\rightarrow u_{\tau,\Gamma}$; and in this context
$u=\{u_{1},\dots,u_{M}\}$ and $u_{\tau}=\{u_{\tau,\Gamma}\}$. We
will use $k=1,\dots,N_{\textrm{site}}$ to label reciprocal lattice
points, $i=1,\dots,N_{\textrm{site}}$ for real space lattice points,
$\alpha=1,\dots,N_{\textrm{orb}}$ for orbitals, and $\sigma=\uparrow,\downarrow$
for spin; and therefore $L=2N_{\textrm{site}}N_{\textrm{orb}}$. Furthermore,
the variational parameters become $\lambda=\{\lambda_{k}\}$, where
$\lambda_{k}=\{\boldsymbol{\lambda}_{1,k},\dots,\boldsymbol{\lambda}_{M+1,k}\}$
and $\boldsymbol{\lambda}_{\tau,k}$ is a matrix of dimension $2N_{\textrm{orb}}\times2N_{\textrm{orb}}$,
which is the $(k,k)$ sub-block of the matrix $\boldsymbol{\lambda}_{\tau}$
. 

The decoupled minimization algorithm for the case of translation symmetry
is summarized as follows. In each iteration, we start with $\lambda,u,\bm{\mathcal{G}}$,
and for any site $i$ we have 
\begin{equation}
\barhat[\rho]_{loc;i}\left(u,\boldsymbol{\mathcal{G}}\right)=\barhat[\rho]_{loc;i}^{0}(\bm{\mathcal{G}})\prod_{\tau}\barhat[P]_{i}^{\left(\tau\right)}\left(u_{\tau}\right).
\end{equation}
We can then compute the local integer time Green's function $\bm{g}_{loc}\left(u,\boldsymbol{\mathcal{G}}\right)=\left\langle \barhat[\boldsymbol{n}]_{i}\right\rangle _{\barhat[\rho]_{loc;i}}$,
the interaction energy $E_{loc}\left(u,\mathcal{G}\right)=N_{\textrm{site}}\langle\barhat[H]_{loc;i}^{\left(\mathcal{N}\right)}\rangle_{\barhat[\rho]_{loc;i}}$,
and the exponential form of the integer time self-energy

\begin{equation}
\bm{S}{}_{loc}\left(u,\boldsymbol{\mathcal{G}}\right)=\left(\bm{\mathcal{G}}^{-1}-\bm{1}\right)^{-1}\left(\bm{g}_{loc}{}^{-1}-\bm{1}\right).
\end{equation}
For each $k$ point, we define 
\begin{align}
\barhat[\rho]_{K,k}\left(\lambda_{k},u,\boldsymbol{\mathcal{G}}\right)= & \barhat[Q]\exp\left(-\sum_{\tau}\ln\left(\bm{\lambda}_{\tau,k}^{-1}-\boldsymbol{1}\right)^{T}\cdot\barhat[\bm{n}]_{k}^{(\tau)}\right)\nonumber \\
 & \times\exp\Big(-\ln\bm{S}_{loc}^{T}\cdot\barhat[\bm{n}]_{k}\Big),
\end{align}
and compute the integer time Green's function $\bm{g}'_{k}\left(\lambda_{k},\boldsymbol{S}_{loc}\right)=\left\langle \barhat[\boldsymbol{n}]_{k}\right\rangle _{\barhat[\rho]_{K,k}},$
the local integer time Green's function $\bm{g}_{loc}'\left(\lambda,u,\boldsymbol{\mathcal{G}}\right)=N_{\textrm{site}}^{-1}\sum_{k}\bm{g}'_{k},$
the kinetic energy $K\left(\lambda,u,\boldsymbol{\mathcal{G}}\right)=\sum_{k}\langle\boldsymbol{\epsilon}_{k}\cdot\barhat[\bm{n}]_{k}^{\left(\discn\right)}\rangle_{\barhat[\rho]_{K,k}}$,
and the constraint $\bm{\Delta}\left(\lambda,u,\boldsymbol{\mathcal{G}}\right)=\bm{g}_{loc}-\bm{g}_{loc}'$.
The iteration procedure to update $\bm{\mathcal{G}}$ becomes 
\begin{equation}
\bm{\mathcal{G}}'=(\boldsymbol{1}+\left(\bm{g}_{loc}'{}^{-1}-1\right)\bm{S}_{loc}^{-1})^{-1}.
\end{equation}
Similarly, $\lambda$ can be updated as 
\begin{equation}
\lambda'_{k}=\text{argmin}_{\lambda_{k}}\langle\boldsymbol{\epsilon}_{k}\cdot\barhat[\bm{n}]_{k}^{\left(\discn\right)}+\text{\ensuremath{\bm{v}}}_{K}\cdot\barhat[\bm{n}]_{k}\rangle_{\barhat[\rho]_{K,k}\left(\lambda_{k},\boldsymbol{S}_{loc}\right)},\label{eq:update_lambda_k}
\end{equation}
where $\bm{v}_{K}=\frac{1}{N_{\textrm{site}}}\frac{\partial E}{\partial\bm{\mathcal{G}}}\cdot\frac{\partial\bm{\mathcal{G}}}{\partial\bm{\Delta}}$.
The update for $u$ simplifies to 

\begin{align}
 & u'=\text{argmin}_{u}\langle\barhat[H]_{loc,i}^{\left(\discn\right)}+\bm{v}_{loc}\cdot\barhat[\bm{n}]_{i}\rangle_{\barhat[\rho]_{loc;i}\left(u,\mathcal{G}\right)},\label{eq:update_u_trans}\\
 & \bm{v}_{loc}=\frac{1}{N_{\textrm{site}}}\left(\frac{\partial K}{\partial\bm{g}_{loc}}-\frac{\partial E}{\partial\bm{\mathcal{G}}}\cdot\frac{\partial\bm{\mathcal{G}}}{\partial\bm{\Delta}}\cdot\frac{\partial\bm{\Delta}}{\partial\bm{g}_{loc}},\right).
\end{align}
For the special case of $\discn=2$, which recovers the usual Gutzwiller
approximation, this decoupled minimization algorithm can be simplified.
First, $[\boldsymbol{\lambda}_{1,k}]_{\alpha\sigma,\alpha'\sigma'}=\frac{1}{2}\delta_{\alpha\alpha'}\delta_{\sigma\sigma'}$
and $\boldsymbol{\lambda}_{2,k}$ is chosen as the single-particle
density matrix of the non-interacting Hamiltonian at $k$. Second,
the self-consistency can be fulfilled \textit{a priori} by choosing
$\bm{\mathcal{G}}=\langle\barhat[\boldsymbol{n}]_{i}\rangle_{\spdcs_{0}\left(\lambda\right)}$
if one enforces $\langle\barhat[\boldsymbol{n}]_{i}^{(2)}\rangle_{\spdcs_{0}\left(\lambda\right)}=\langle\barhat[\boldsymbol{n}]_{i}^{(2)}\rangle_{\barhat[\rho]_{loc}}$
\cite{Cheng2021195138}. In this case, only $u$ must be updated during
each iteration. Alternatively, both $\boldsymbol{\mathcal{G}}$ and
$u$ would need to be updated.

\section{The Two Band Hubbard model with $\discn=2,3,4$}

\subsection{Hamiltonian for the two band Hubbard model and the SPD\label{subsec:Hamiltonian-for-the}}

In this paper, we focus on the two orbital Hubbard model on the Bethe
lattice in $d=\infty$. The local portion of the Hamiltonian consists
of the crystal field splitting and the Slater-Kanamori parameterization
of the local interaction, given as 
\begin{align}
\hat{H}_{loc;i}= & \Delta\sum_{\sigma}(\hat{n}_{i1\sigma}-\hat{n}_{i2\sigma})-\mu\sum_{\alpha\sigma}\hat{n}_{i\alpha\sigma}\nonumber \\
 & +U\hat{O}_{i1}+U'\hat{O}_{i2}+\left(U'-J\right)\hat{O}_{i3}-J\hat{O}_{i4}
\end{align}
where
\begin{align}
 & \hat{O}_{i1}=\sum_{\alpha=1,2}\hat{n}_{i\alpha\uparrow}\hat{n}_{i\alpha\downarrow},\label{eq:densdens1}\\
 & \hat{O}_{i2}=\sum_{\sigma}\hat{n}_{i1\sigma}\hat{n}_{i2\bar{\sigma}},\hspace{1em}\hat{O}_{i3}=\sum_{\sigma}\hat{n}_{i1\sigma}\hat{n}_{i2\sigma},\label{eq:densdens2}\\
 & \hat{O}_{i4}=\left(\hat{a}_{i1\downarrow}^{\dagger}\hat{a}_{i1\uparrow}\hat{a}_{i2\uparrow}^{\dagger}\hat{a}_{i2\downarrow}+\hat{a}_{i1\downarrow}^{\dagger}\hat{a}_{i2\uparrow}\hat{a}_{i1\uparrow}^{\dagger}\hat{a}_{i2\downarrow}+h.c.\right),\label{eq:densdens4}
\end{align}
where $\Delta$ is the crystal field, $U$ and $U'=U-2J$ are on-site
intraorbital and interorbital Coulomb interactions, respectively,
and $J$ is the Hund coupling (see \cite{supplementary} for eigenvalues
and eigenvectors of the local Hamiltonian). 

It is instructive to deduce the limiting behavior of $U$, $J$, and
$\Delta$ for a given hopping parameter $t$, some of which has been
discussed previously \cite{Werner2007126405}. For small values of
$U$ and $J$ where the system is not strongly polarized (i.e. $n_{1\sigma}\not\rightarrow0$),
the susceptibility $\partial n_{1\sigma}/\partial\Delta$ is dictated
by the non-interacting Hamiltonian. For large $U$ and small $\Delta$,
the system is insulating and $J/\Delta$ will determine the nature
of the insulator. There will be a competition between the spin triplet
state with energy $U-3J$ and the spin singlet state with energy $U-\sqrt{4\Delta^{2}+J^{2}}$,
and a transition will occur for $\Delta_{c}=\sqrt{2}J$. For $\Delta<\Delta_{c}$,
the system will be in the triplet state and $n_{\alpha\sigma}=\frac{1}{2}$,
and for $\Delta>\Delta_{c}$ the system will be in the singlet state
where $n_{1\sigma}=(1/2-\Delta/\sqrt{4\Delta^{2}+J^{2}})$. For $\Delta=\Delta_{c}$,
the singlet state will have $n_{1\sigma}^{\star}=1/2-\sqrt{2}/3\approx0.0286$.
For small $J/\Delta$, we have $n_{1\sigma}=J^{2}/(16\Delta^{2})+\dots$.
Finally, for small $U$ and $J$ where the system is strongly polarized
(i.e. $n_{1\sigma}\rightarrow0$), there will be a competition between
the kinetic energy and the local interactions. The kinetic energy
will scale like $tn_{1\sigma}$ and the dominant interaction energy
will scale like $J\sqrt{n_{1\sigma}}$, and therefore a metal-insulator
transition (MIT) phase boundary $n_{1\sigma}\propto J^{2}/t^{2}$
should be anticipated.

The interacting projector of the SPD (Eq. \ref{eq:int_proj_i}) is
defined using $\hat{P}_{i\Gamma}$, with $\Gamma=1,\dots,18$, and
the first 16 are 
\begin{equation}
\hat{P}_{i\Gamma}=\prod_{\alpha\sigma}\left(\Gamma_{\alpha\sigma}\hat{n}_{\alpha\sigma}+\left(1-\Gamma_{\alpha\sigma}\right)\left(1-\hat{n}_{\alpha\sigma}\right)\right),
\end{equation}
where $\Gamma_{\alpha\sigma}\in\{0,1\}$ and are determined from the
binary relation $\left(\Gamma_{1\uparrow}\Gamma_{1\downarrow}\Gamma_{2\uparrow}\Gamma_{2\downarrow}\right)_{2}=\Gamma-1$
(see \cite{supplementary} for explicit expressions). The remaining
two operators are given as 

\begin{align}
\hat{P}_{i17} & =\hat{a}_{i1\uparrow}^{\dagger}\hat{a}_{i1\downarrow}\hat{a}_{i2\downarrow}^{\dagger}\hat{a}_{i2\uparrow}+h.c.,\\
\hat{P}_{i18} & =\hat{a}_{i1\uparrow}^{\dagger}\hat{a}_{i2\downarrow}\hat{a}_{i1\downarrow}^{\dagger}\hat{a}_{i2\uparrow}+h.c.
\end{align}
For the non-interacting projector, we use $N_{\textrm{site}}=40$,
which proved to be sufficiently converged.

\begin{figure}[tbph]
\includegraphics[width=1\columnwidth]{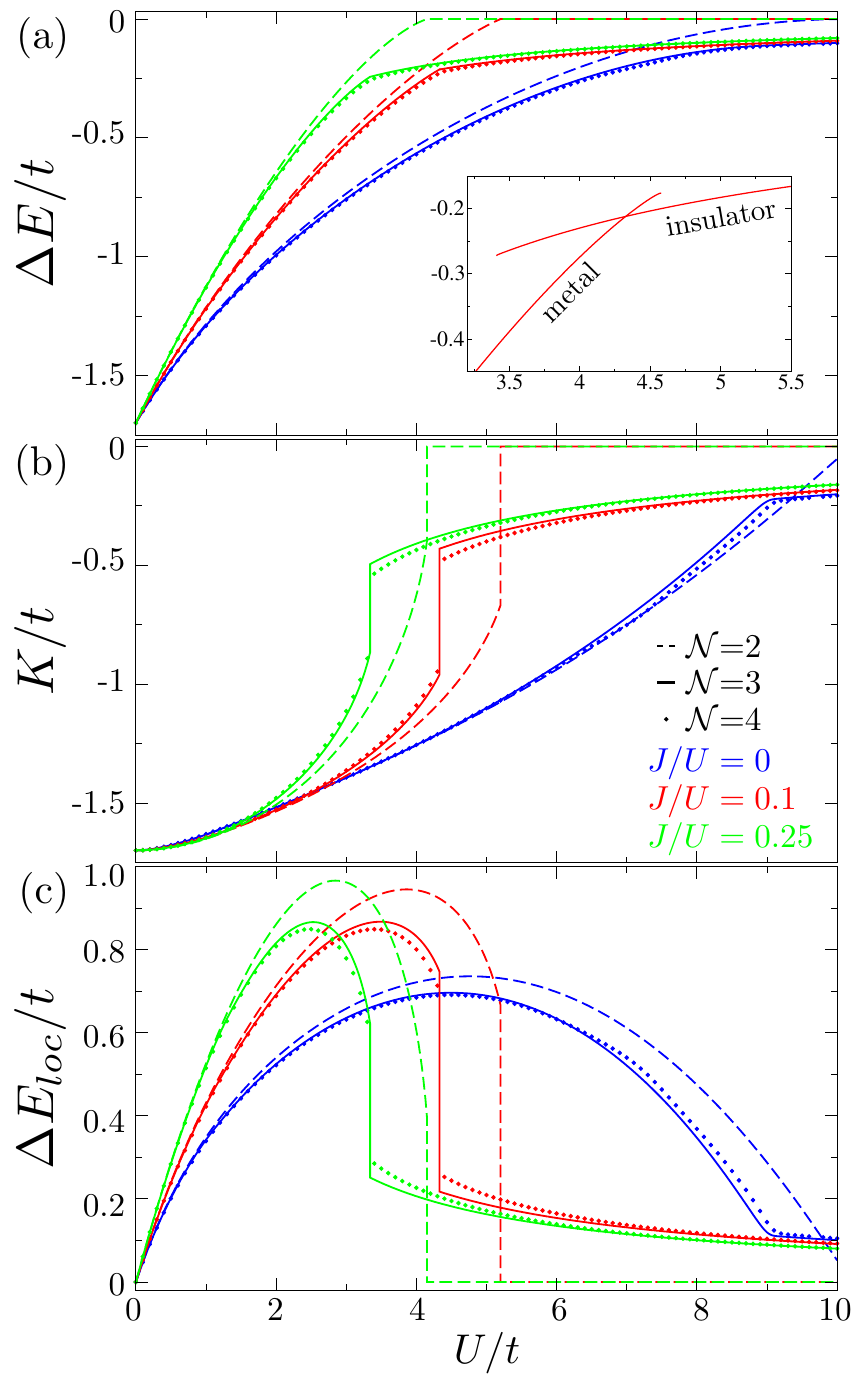}

\caption{\label{fig:total-energy} Zero temperature energetics as a function
of $U/t$ for various $J/U$ at $\Delta=0$ in the two band Hubbard
model for the $d=\infty$ Bethe lattice at half filling. VDAT results
for $\discn=2,3,4$ are provided. ($a$) The total energy difference
$\Delta E(t,U,J)=E(t,U,J)-E(0,U,J)$, where $E(t,U,J)$ is the total
energy per site. Inset shows metastable region for $\discn=3$ at
$J/U=0.1$. ($b$) Kinetic energy per site $K=\langle\hat{K}\rangle/N_{\textrm{site}}$.
($c$) Interaction energy difference $\Delta E_{loc}(t,U,J)=E_{loc}(t,U,J)-E_{loc}(0,U,J)$,
where $E_{loc}(t,U,J)$ is the interaction energy per site.}
\end{figure}

\subsection{Computational complexity of the SCDA}

The power of VDAT within the SCDA is that it can be exactly evaluated
for $d=\infty$ at a very small computational cost, and here we examine
the computational complexity. For simplicity, we focus on the case
with translation symmetry, outlined in Section \ref{subsec:Translation-Symmetry}.
Within a given iteration of the SCDA, there are two relevant scalings
to consider: the aspects relevant to $\barhat[\rho]_{K,k}\left(\lambda,\boldsymbol{S}_{loc}\right)$
and $\barhat[\rho]_{loc}\left(u,\boldsymbol{\mathcal{G}}\right)$.
For the former, the complexity scales linearly in $N_{\textrm{site}}$
and polynomially in $N_{\textrm{orb}}\mathcal{N}$. For the latter,
the complexity is independent of $N_{\textrm{site}}$ and scales exponentially
in $N_{\textrm{orb}}\mathcal{N}$. Normally, the exponential scaling
will be the limiting factor, and therefore we focus on showcasing
the cost in specific examples for the two orbital Hubbard model. 

For a given $N_{\textrm{orb}}$ and $\mathcal{N}$, there are three
relevant tasks for evaluating expectation values under $\barhat[\rho]_{loc}\left(u,\boldsymbol{\mathcal{G}}\right)$.
First, given an input $u$ and $\boldsymbol{\mathcal{G}}$, the $\boldsymbol{g}_{loc}$
and $E_{loc}$ must be computed. Second, the $\boldsymbol{v}_{K}$
and $\boldsymbol{v}_{loc}$ require the computation of the first derivatives
$\partial\boldsymbol{g}_{loc}/\partial\boldsymbol{\mathcal{G}}$ and
$\partial E_{loc}/\partial\boldsymbol{\mathcal{G}}$ . Third, Eq.
\ref{eq:update_u_trans} must be minimized with respect to $u$, requiring
the computation of the polynomial coefficients of $\boldsymbol{g}_{loc}$
and $E_{loc}$ in $u$ given in Eq. \ref{eq:local_operator_expansion_u}.
For each task, the compiled machine code size and the execution time
is provided for $\discn=2,3,4$ in Table \ref{tab:Computational-cost-for}.
The machine code size is proportional to the number of instructions
which need to be executed. We provide the execution time for $N_{\textrm{orb}}=2$
using a single processor core, demonstrating that the compute time
is approximately proportional to the machine code size. The tiny computational
times on this modest computational resource illustrates the power
of VDAT for the two-orbital Hubbard model. Indeed, all VDAT results
generated in this study were executed using a single processing core.
It should be noted that $\discn=2$ and $\discn=3$ are on the same
scale, which is true for all $\discn=2M$ and $\discn=2M+1$ for $M>0$,
and this can be understood from the fact that these two cases share
the same number of interacting projectors. Given that the minimization
over all variational parameters can be achieved on the order of 10
iterations, the total computation time for $\discn=4$ is roughly
estimated by the cost of task 2 times 10, yielding $\approx5$ seconds.
The total computation time for $\discn\le3$ is less than one second,
and it is dominated by aspects relevant to $\barhat[\rho]_{K,k}\left(\lambda,\boldsymbol{S}_{loc}\right)$.
\begin{table}
\begin{tabular}{|c|l|l|l|l|l|l|}
\hline 
 & \multicolumn{2}{c|}{Task1} & \multicolumn{2}{c|}{Task 2} & \multicolumn{2}{c|}{Task 3}\tabularnewline
\hline 
$\discn$ & time (s) & size (Mb) & time (s) & size (Mb) & time (s) & size (Mb)\tabularnewline
\hline 
\hline 
2 & $2\cdot10^{-5}$ & $1.3\cdot10^{-1}$ & $6\cdot10^{-5}$ & $5.1\cdot10^{-1}$ & $4\cdot10^{-5}$ & $2.7\cdot10^{-1}$\tabularnewline
\hline 
3 & $7\cdot10^{-5}$ & $5.2\cdot10^{-1}$ & $3\cdot10^{-4}$ & $3.3$ & $2\cdot10^{-4}$ & $1.9$\tabularnewline
\hline 
4 & $4\cdot10^{-2}$ & $2.6\cdot10^{2}$ & $5\cdot10^{-1}$ & $3.0\cdot10^{3}$ & $8\cdot10^{-2}$ & $7.2\cdot10^{2}$\tabularnewline
\hline 
\end{tabular}

\caption{\label{tab:Computational-cost-for}Computational cost for various
tasks within the SCDA for the two orbital Hubbard model at different
$\discn$. Each of the three tasks is defined in the main text. The
compiled machine code size of the corresponding function and the execution
time on a single processor core is provided.}
\end{table}

To better understand the scaling for aspects relevant to $\barhat[\rho]_{loc}\left(u,\boldsymbol{\mathcal{G}}\right)$,
it useful to think in terms of the number of effective orbitals $N_{\textrm{eff}}$,
which is the number of spin orbitals in the compound space that have
a nontrivial interacting projector. For example, for $N_{\textrm{orb}}=2$
at $\discn=2-3$ we have $N_{\textrm{eff}}=8$, while for $\discn=4-5$
we have $N_{\textrm{eff}}=16$. The computational time $t$ for a
given task scales exponentially with $N_{\textrm{eff}}$, so we approximately
have $t=c_{0}c_{1}^{N_{\textrm{eff}}}$, and using the results from
Table \ref{tab:Computational-cost-for} for task 1 with $\discn=2$
and $\discn=4$, we estimate $c_{0}=10^{-8}$ seconds and $c_{1}=2.6$.
Using this simple parametrization, we can estimate the time required
for $N_{\textrm{orb}}=5$ and $\discn=2-3$, where $N_{\textrm{eff}}=20$,
resulting in $t=1.8$ seconds on a single core. In the absence of
any symmetry, there will be on the order of $2^{10}$ variational
parameters which will need to be minimized. Overall, it appears that
generally treating $d$-orbitals with the present algorithm should
be tractable for $\discn\le3$. We can also estimate the time required
for $N_{\textrm{orb}}=7$ and $\discn=2-3$, where $N_{\textrm{eff}}=28$,
resulting in $t=3.6\times10^{3}$ seconds on a single core. In the
absence of symmetry, there will be on the order of $2^{14}$ variational
parameters. Therefore, it appears that generally treating $f$-electrons
will require parellelization, which can be achieved in a number of
ways. Perhaps the simplest approach would be to perform a generalized
Hubbard-Stratonovich transformation, recasting the interacting projectors
into a sum of non-interacting projectors, and allowing the evaluation
of Eq. \ref{eq:local_operator_expansion_u} in a perfectly parallel
fashion; dividing the task 1 cost of $t=3.6\times10^{3}$ by the number
of available cores. Therefore, generally treating $f$-electrons for
$\discn\le3$ using reasonable computational resources appears completely
tractable. 

\subsection{Results\label{sec:Results-for-Two}}

We now illustrate VDAT for the two-orbital Hubbard model in $d=\infty$,
where the SCDA exactly evaluates the SPD. Our VDAT results stand alone
in the sense that the results monotonically approach the exact solution
as $\discn$ increases. However, we also compare to published DMFT
results using the CTQMC algorithm to solve the DMFT impurity problem,
which recovers the numerically exact results at a finite temperature
\cite{Werner2007126405}. Given that our VDAT results are at zero
temperature, one must compare to the finite temperature DMFT results
with caution, as the insulating regime will be rather sensitive to
temperature. We will focus on the half filled case of two electrons
per site in the paramagnetic state (i.e. $n_{1\sigma}+n_{2\sigma}=1$).

\begin{figure}[tbph]
\includegraphics[width=0.98\columnwidth]{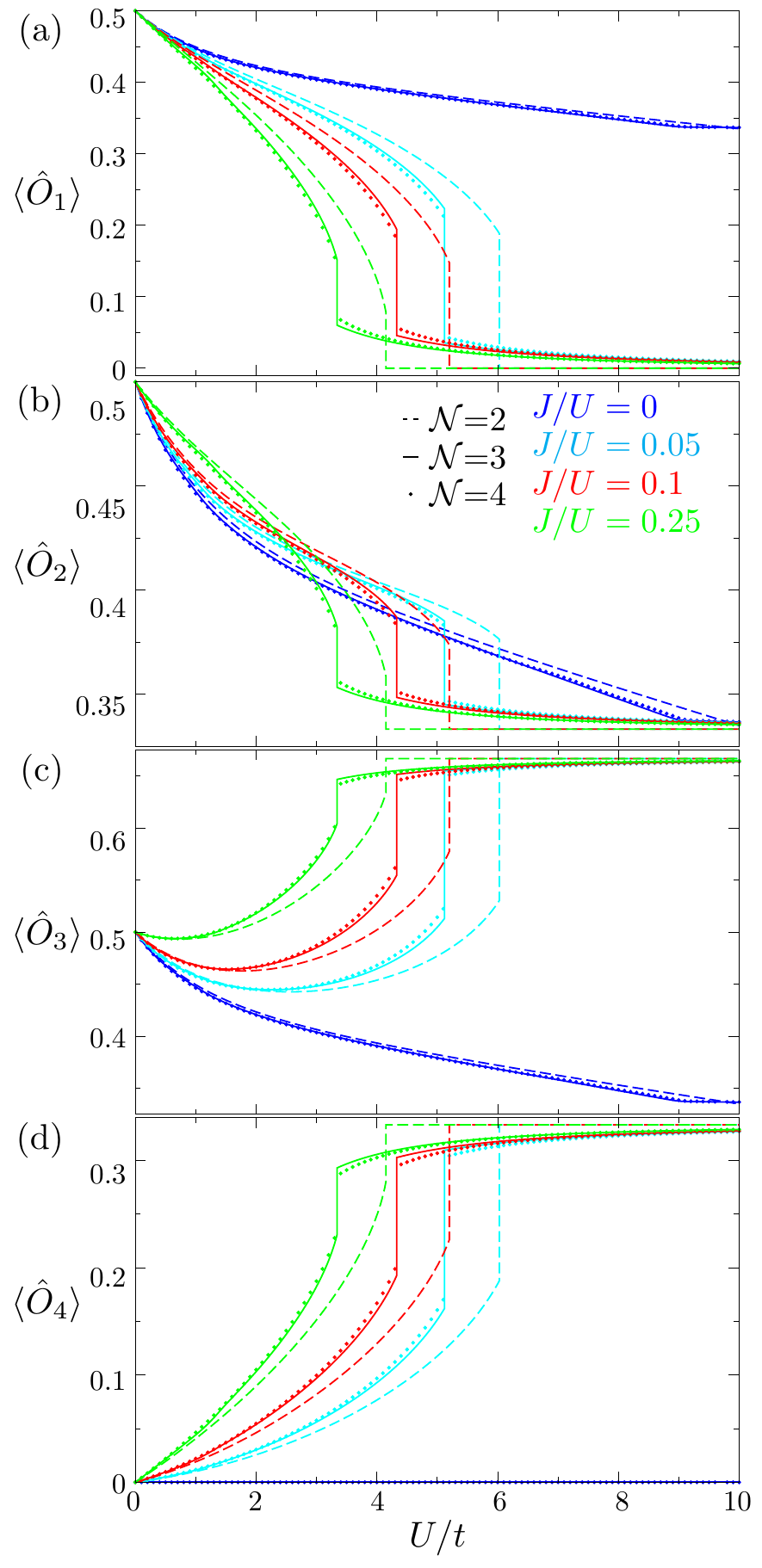}

\caption{\label{fig:Interaction-energy.}Local two particle correlation functions
as a function of $U/t$ for various $J/U$ at $\Delta=0$ in the two
band Hubbard model for the $d=\infty$ Bethe lattice at half filling.
VDAT results for $\discn=2,3,4$ are provided. Panels $a$-$d$ are
expectation values of the operators defined in Eqns. \ref{eq:densdens1}-\ref{eq:densdens4}.}
\end{figure}

We begin by considering the all of the different components of the
energy at $\Delta=0$ for a broad range of $J$ and $U$. The total
energy is computed using $\discn=2,3,4$ with $J/U=0,0.1,0.25$ over
a dense grid of $U/t\in[0,10]$ (see Figure \ref{fig:total-energy},
panel $a$). In order to facilitate comparison, we plot the difference
in the total energy and the atomic energy, where the latter is $E(0,U,J)=U-3J$.
For a given $U$ and $J$, the total energy strictly decreases as
$\discn$ increases, as is required by the variational principle.
The $\discn=2$ result recovers the usual Gutzwiller approximation,
and yields an insulator which is simply a collection of atoms. Clearly,
$\discn=3$ produces a substantial quantitative improvement over $\discn=2$,
in addition to a realistic insulating state which allows for virtual
hopping. The $\discn=4$ result only produces a small quantitative
change as compared to $\discn=3$, demonstrating that both $\discn=3$
and $\discn=4$ are close to the exact solution. We now focus on the
qualitative nature of the MIT. For $J/U>0$, a clear kink in the total
energy as a function of $U/t$ can be observed for all $\discn$,
indicating a first-order MIT. For $\discn=3$, we illustrate the metastable
regime by initiating calculations from both metallic and insulating
solutions (see Figure \ref{fig:total-energy}, panel $a$ inset).
Alternatively, for $J/U=0$, the MIT is continuous for all $\discn$.
For $\discn=2$, our results are consistent with previous findings
using the Gutzwiller approximation \cite{Bunemann19986896}. For $\discn\ge3$
and $J/U>0$, the fact that the first order transition survives is
consistent with DMFT calculations which used DMRG \cite{Hallberg201517001}
or NRG\cite{Pruschke2005217} solvers. 

It is also interesting to separately consider the kinetic and interaction
energy (see Figure \ref{fig:total-energy}, panel $b$ and $c$, respectively),
which probe the derivative of the total energy with respect to $t$
and $U$ (assuming fixed $J/U$), respectively. The kinetic energy
increases with increasing $\discn$ in the metallic regime and decreases
in the insulating regime. The opposite behavior is observed for the
interaction energy. Additionally, a clear discontinuity can be observed
at the MIT for $J/U>0$, as the MIT is first-order, whereas a kink
is observed for $J/U=0$ as the MIT is continuous. 

In order to understand the competition between $U$ and $J$, it is
useful to study the individual components of the interaction energy,
defined in Eqns. \ref{eq:densdens1}-\ref{eq:densdens4}, and we plot
each as a function of $U/t$ for various $J/U$ (see Figure \ref{fig:Interaction-energy.}).
The $\langle\hat{O}_{1}\rangle$, $\langle\hat{O}_{2}\rangle$, and
$\langle\hat{O}_{4}\rangle$ all change monotonically as a function
of $U$, dictated by the sign of the respective coupling coefficients
in the local Hamiltonian, while $\langle\hat{O}_{3}\rangle$ is nonmonotonic
in $U$ for finite $J/U$. In the small $U$ regime, the $\langle\hat{O}_{3}\rangle$
decreases, commensurate with the respective coupling coefficient,
while in the large $U$ regime it increases due to the prominence
of the triplet states. 
\begin{figure}[tbph]
\includegraphics[width=0.95\columnwidth]{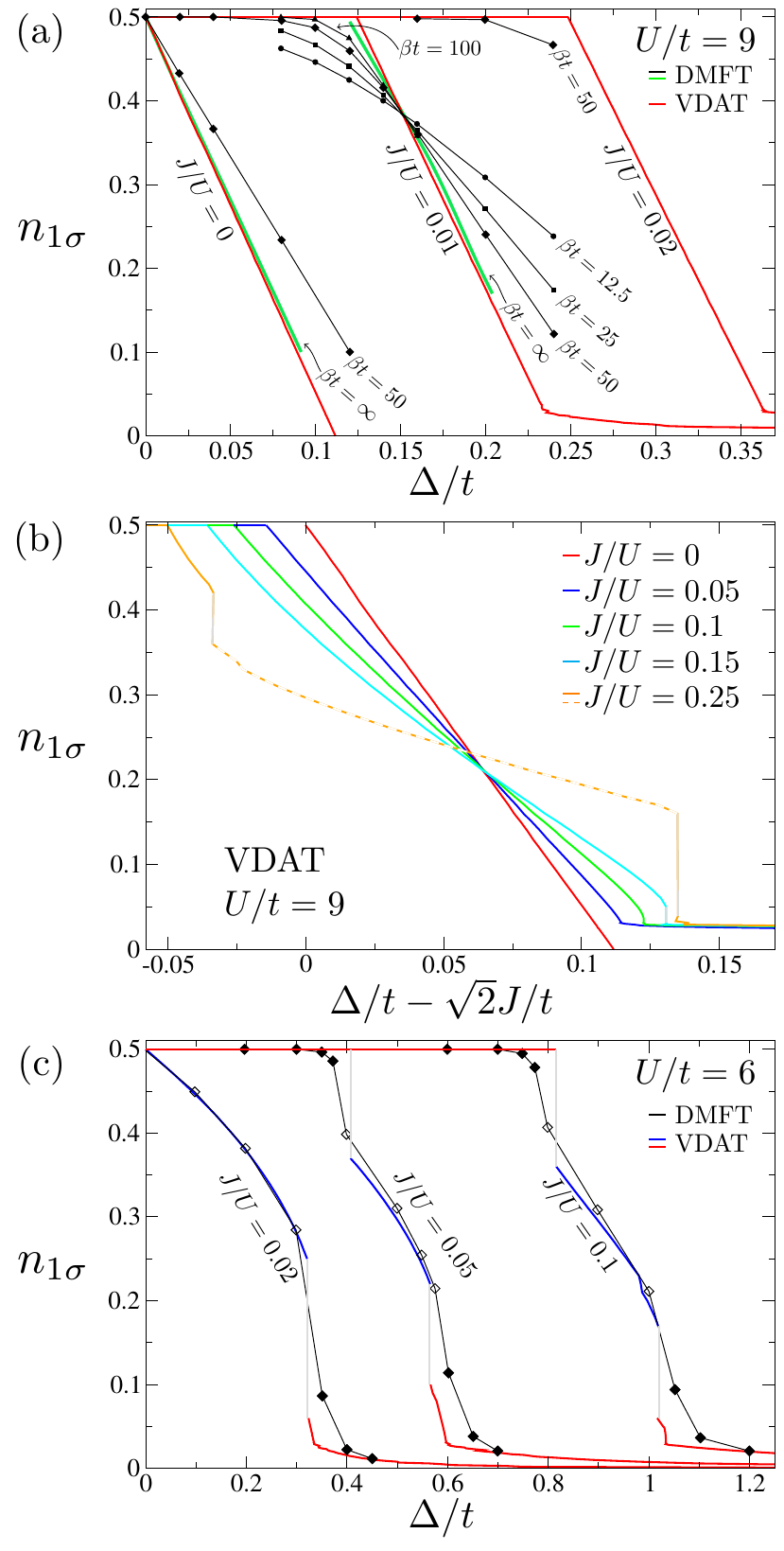}

\caption{\label{fig:n1_vs_Delta_fix_U_J}The occupancy $n_{1\sigma}$ as a
function of $\Delta$ for various $J/U$ at a given $U/t$ in the
two band Hubbard model for the $d=\infty$ Bethe lattice at half filling.
VDAT results are presented for $\discn=3$, in addition to published
DMFT (CTQMC) results\cite{Werner2007126405} at various temperatures.
($a$) Results for $U/t=9$; all VDAT and DMFT results are insulators;
thick green lines are extrapolations of the DMFT results to zero temperature
(see text for details). ($b$) VDAT results for $U/t=9$ at larger
values of $J/U$, where the $x$-axis is shifted by $\sqrt{2}J/t$.
Solid lines are insulators, while the dashed line is a metal. ($c$)
Results for $U/t=6$. Solid (hollow) DMFT points are insulating (metallic),
and red (blue) VDAT lines are insulating (metallic).}
\end{figure}

We now proceed to more thoroughly explore nonzero $\Delta$, and we
begin by examining $n_{1\sigma}$ as a function of $\Delta/t$ at
relatively large values of $U/t=6,9$ for various $J/U$ (see Figure
\ref{fig:n1_vs_Delta_fix_U_J}), which can be compared to previous
DMFT calculations \cite{Werner2007126405}. We begin by making general
observations about the large $U/t$ insulating regime, where the case
of $J=0$ and $J>0$ are qualitatively different (see Figure \ref{fig:n1_vs_Delta_fix_U_J},
panel $a$). For $J/U=0$, the system has an approximately constant
orbital susceptibility for $\Delta<\Delta_{c}$ and is fully polarized
for $\Delta>\Delta_{c}$, where $\Delta_{c}\approx t^{2}/U$. We refer
to these two regions as partially and fully orbitally polarized insulators,
respectively. For $J/U>0$, the orbital susceptibility is zero for
$\Delta<\Delta_{c_{1}}$, approximately constant for $\Delta_{c_{1}}<\Delta<\Delta_{c_{2}}$,
and determined from $n_{1\sigma}\approx J^{2}/(16\Delta^{2})$ for
$\Delta>\Delta_{c_{2}}$; where $\Delta_{c_{1}}\approx\sqrt{2}J$
and $\Delta_{c_{2}}\approx\Delta_{c_{1}}+t^{2}/U$. We refer to these
three regions as zero, partially, and largely orbitally polarized
Mott insulators. The underlying physics of these different regions
has been discussed previously \cite{Werner2007126405}, but the nature
of the transition between these regions at zero temperature has not
been resolved. 

At $U/t=9$, the system is insulating for all values of $\Delta/t$
and $J/U$ (see Figure \ref{fig:n1_vs_Delta_fix_U_J}, panel $a$).
Interestingly, the polarization (i.e. $n_{2\sigma}-n_{1\sigma}$)
at the transition point between the partially and largely orbitally
polarized insulators is roughly independent of $J$ and $U$, and
occurs at $n_{1\sigma}\approx n_{1\sigma}^{\star}=1/2-\sqrt{2}/3\approx0.0286$.
In contrast to $J=0$, for finite $J$ the system will not fully polarize
for finite $\Delta$. The previously published DMFT results are strongly
affected by temperature, as illustrated by the calculations for $J/U=0.01$
at $\beta t=12.5,25,50,100$. We extrapolate the DMFT results to $\beta t\rightarrow\infty$,
which agrees well with our zero temperature $\discn=3$ results for
the zero and partially orbitally polarized insulators, while the DMFT
results were not computed for the largely orbitally polarized insulator.
For $J/U=0$, only $\beta t=50$ was computed with DMFT, and we approximately
extrapolate their results to zero temperature by approximating the
entropy. The crystal field for a given $n_{1\sigma}$ can be computed
from the free energy as $\Delta_{T}=-\left(1/4\right)\partial F(T,n_{1\sigma})/\partial n_{1\sigma}$,
where $F(T,n_{1\sigma})=E(T,n_{1\sigma})-TS(T,n_{1\sigma})$ and $E(T,n_{1\sigma})$
is the Legendre transform of the total energy with respect to $\Delta$
where the polarization is parametrized in terms of $n_{1\sigma}$.
We assume that $E(0,n_{1\sigma})\approx E(T,n_{1\sigma})$ for small
$T$, so the zero temperature crystal field can be approximated as
$\Delta_{T=0}=-\left(1/4\right)\partial E(T,n_{1\sigma})/\partial n_{1\sigma}$.
To estimate $S(T,n_{1\sigma})$, we use the atomic limit to approximate
$S(T,1/2)\approx\ln6$ and $S(T,0)=0$. Given the symmetry between
$n_{1\sigma}$ and $n_{2\sigma}$, we assume that $S(T,n_{1\sigma})$
is quadratic about $n_{1\sigma}=1/2$, and thus we approximate $S(T,n_{1\sigma})\approx(1-4(n_{1\sigma}-\frac{1}{2})^{2})\ln6$,
yielding $\Delta_{T=0}-\Delta_{T}\approx2\ln6\left(n_{1\sigma}-\frac{1}{2}\right)T.$
The green curve removes this finite temperature contribution from
the DMFT result, yielding excellent agreement with our $\discn=3$
result. For $J=0$, the temperature effect is straightforward: for
$n_{1\sigma}<1/2$, we have $\Delta_{T=0}<\Delta_{T}$ due to the
fact that finite temperature favors the high entropy state with zero
polarization. However, for $J>0$, the effect of temperature is clearly
more subtle. We see temperature plays opposite roles in the small
and large orbital polarization regime. It is also interesting to explore
larger values of $J/U$ (see Figure \ref{fig:n1_vs_Delta_fix_U_J},
panel $b$), and for a convenient comparison, we shift the $x$-axis
by $\sqrt{2}J/t$. As $J/U$ increases, the orbital susceptibility
decreases, and the transition from the partially to largely orbitally
polarized Mott insulator is first order for $J/U=0.1,0.15$. For $J/U=0.25$
the crystal field drives a first-order MIT from a partially orbitally
polarized Mott insulator to a metal followed by another first-order
MIT to a largely orbitally polarized Mott insulator. At $U/t=6$ (see
Figure \ref{fig:n1_vs_Delta_fix_U_J}, panel $c$), the results contain
both metallic and insulating phases, and VDAT can faithfully capture
the details of the metal-insulator transition. The differences between
VDAT and DMFT are relatively small in this case, and are likely attributable
to the finite temperature of the DMFT calculations. 
\begin{figure}[tbph]
\includegraphics[width=0.98\columnwidth]{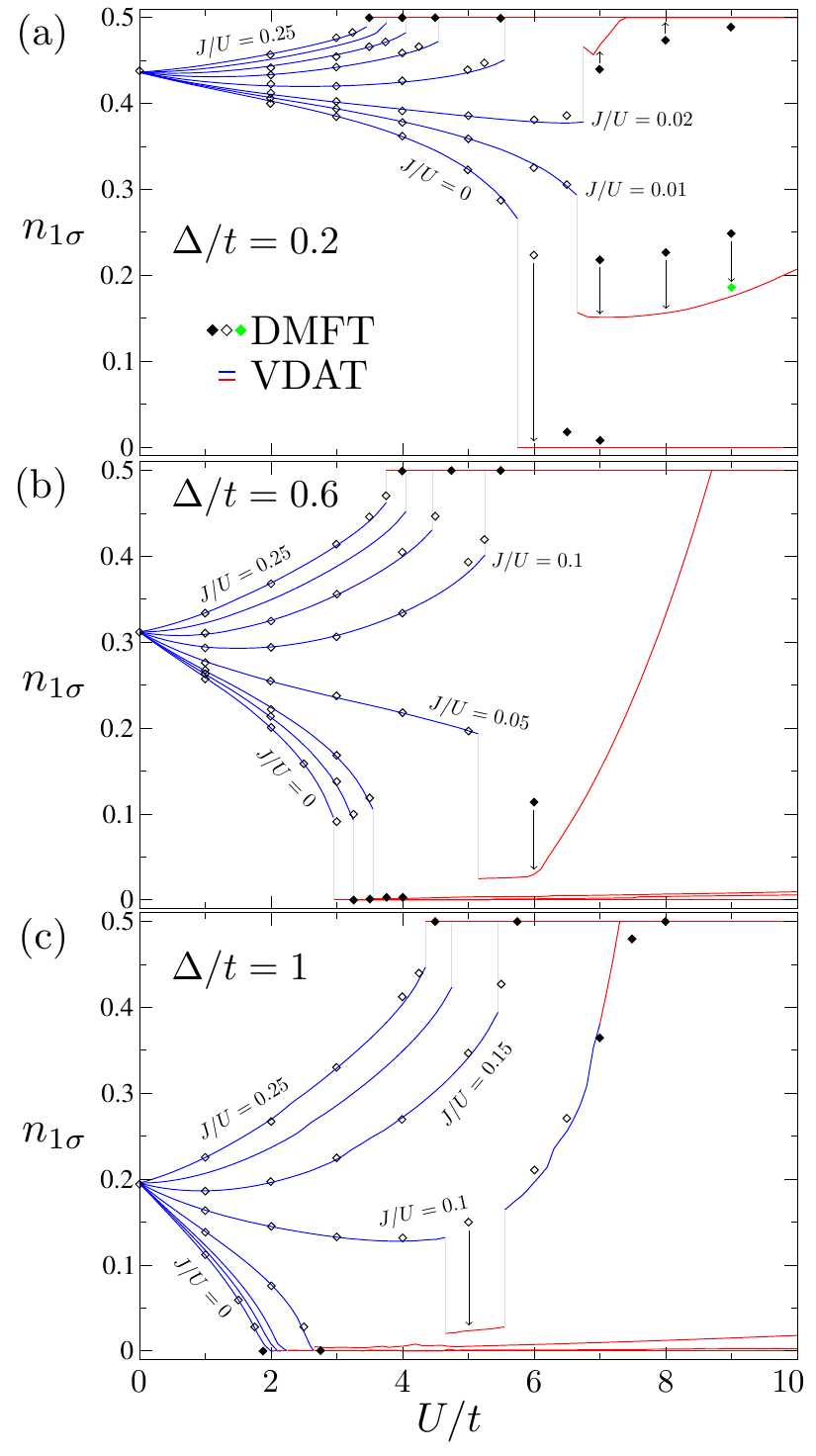}

\caption{\label{fig:n1_vs_U_fix_Delta_J}The occupancy $n_{1\sigma}$ as a
function of $U/t$ for various $J/U$ at a given $\Delta/t$ in the
two band Hubbard model for the $d=\infty$ Bethe lattice at half filling.
VDAT results are presented for $\discn=3$, in addition to published
DMFT (CTQMC) results at $\beta t=50$\cite{Werner2007126405}. VDAT
results for $J/U=0,0.01,0.02,0.05,0.1,0.15,0.2,0.25$ are provided
in each panel (select curves are labeled), and corresponding DMFT
results are provided where available (arrows are used when the correspondence
is unclear). Red (blue) VDAT lines are insulating (metallic), while
black solid (hollow) DMFT points are insulating (metallic); and the
green DMFT point is an extrapolation to zero temperature (see text
for details). ($a$) $\Delta/t=0.2$ ($b$) $\Delta/t=0.6$ ($c$)
$\Delta/t=1$. }
\end{figure}

The previous results focused more on the large $U/t$ regime, and
here we explore a broad range of $U/t$ for various values of $J/U$
and $\Delta/t$ (see Figure \ref{fig:n1_vs_U_fix_Delta_J}). Overall,
there is excellent agreement with DMFT in the metallic region, while
there are nontrivial differences in the insulating regime which are
likely due to the finite temperature of the DMFT calculations. The
zero temperature extrapolation of the DMFT results for $U/t=9$ and
$J/U=0.01$ and $\Delta/t=0.2$ from Figure \ref{fig:n1_vs_Delta_fix_U_J}$a$
is plotted as a green point in Figure \ref{fig:n1_vs_U_fix_Delta_J}$a$,
showing good agreement with our VDAT results. Interestingly, at $\Delta/t=0.6$
and $J/U=0.05$ (see Figure \ref{fig:n1_vs_U_fix_Delta_J}, panel
$b$), increasing $U/t$ drives a transition from a metal to a largely
orbitally polarized Mott insulator which then transitions to a partially
orbitally polarized Mott insulator followed by a zero orbitally polarized
Mott insulator. Similar behavior is observed for $\Delta/t=1.0$ and
$J/U=0.1$ (see Figure \ref{fig:n1_vs_U_fix_Delta_J}, panel $c$),
though the system becomes has an additional first order transition
from the largely orbitally polarized Mott insulator to a metal before
transitioning to the partially orbitally polarized Mott insulator.

\begin{figure*}[tp]
\includegraphics[width=0.98\textwidth]{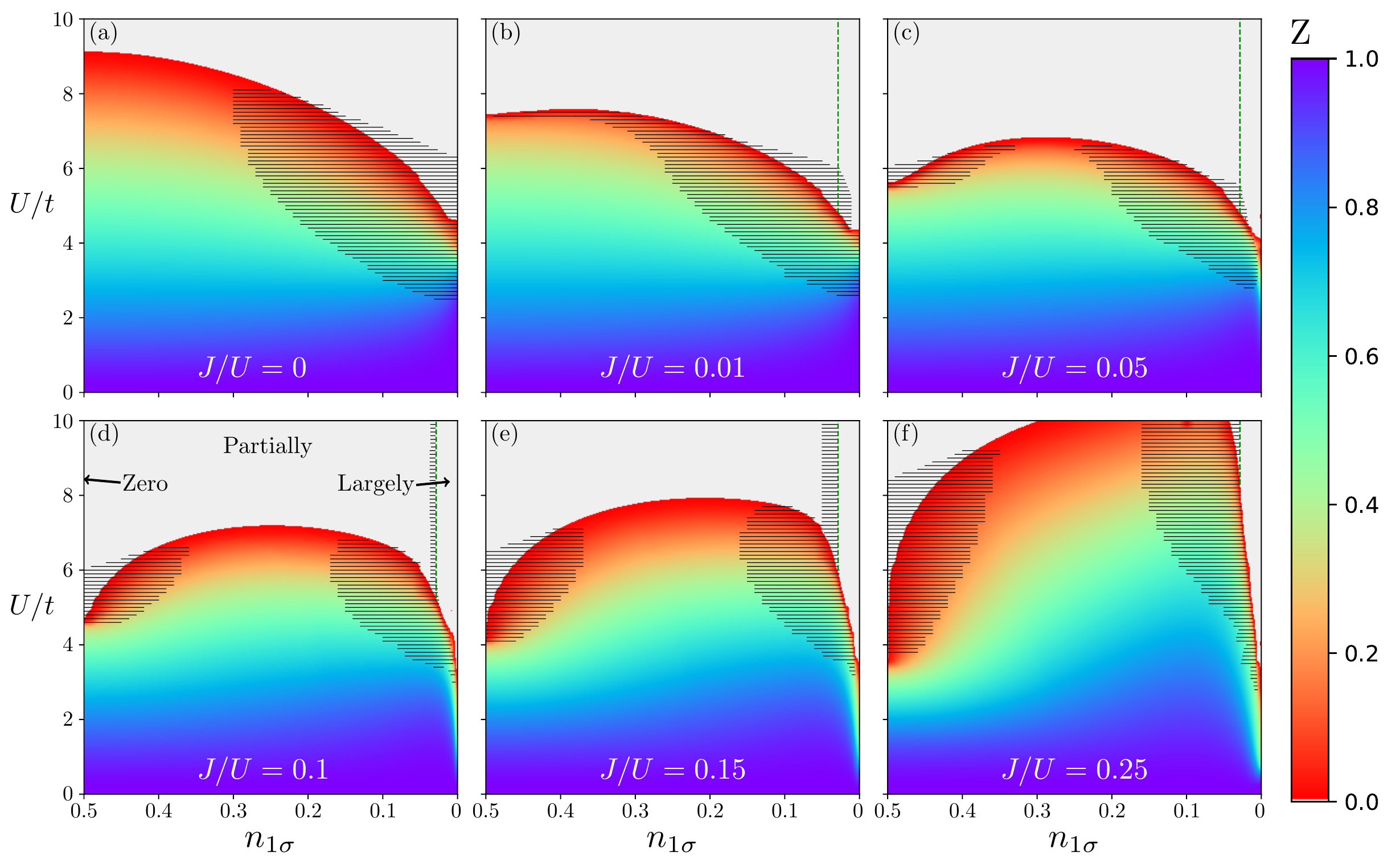}

\caption{\label{fig:phase_diagram} VDAT $\discn=3$ results for the quasiparticle
weight $Z$ as a function of $n_{1\sigma}$ and $U/t$ for various
$J/U$ in the two band Hubbard model for the $d=\infty$ Bethe lattice
at half filling. The quasiparticle weight is zero when the color is
grey. The black hatched lines indicate unstable regions. The vertical
green lines at $n_{1\sigma}=1/2-\sqrt{2}/3\approx0.0286$ indicate
the boundary between the partially and largely orbitally polarized
Mott insulator for $U/t\rightarrow\infty$. For $J/U=0.1$, the zero,
partially, and largely orbitally polarized Mott insulating regimes
are labeled.}
\end{figure*}

In order to obtain a detailed understanding over the entire phase
space of parameters, we evaluate the quasiparticle weight as function
of $U/t$ and $n_{1\sigma}$ for various $J/U$, which serves as a
phase diagram of the metal-insulator transition (see Figure \ref{fig:phase_diagram}).
Additionally, a Maxwell construction is used to determine if a given
value of $n_{1\sigma}$ is stable with respect to $\Delta$, and hatched
lines are used to denote unstable regions. We used two resolutions
for the phase diagram: 0.1 in $U/t$ and 0.01 in $n_{1\sigma}$ for
$n_{1\sigma}>0.05$, and 0.01 in $U/t$ and 0.001 in $n_{1\sigma}$
for $n_{1\sigma}<0.05$, yielding a total of 55,000 calculations per
phase diagram, and this level of resolution would be formidable using
DMFT. In the region of large polarization (i.e. $n_{1\sigma}\rightarrow0$),
the kinetic energy is approaching zero, making the convergence of
the calculation challenging; this region would be best explored by
treating $n_{1\sigma}$ as a perturbation parameter, but we leave
this for future work. For $J/U=0$ (see Figure \ref{fig:phase_diagram},
panel $a$), at $n_{1\sigma}=1/2$ there is a MIT at $U/t=9.1$, and
the transition value of $U/t$ decreases monotonically for decreasing
$n_{1\sigma}$. In the large polarization limit where $n_{1\sigma}\rightarrow0$,
there is a band-insulator to fully polarized Mott insulator transition
at finite $U/t$. Furthermore, for $2.3\le U/t\le8$ there is a first-order
MIT driven by $\Delta$ (denoted by hatching), and three regimes can
be seen. For the smallest $U/t$ region, there is a metal to band
insulator transition; for intermediate $U/t$, there is a metal to
fully orbitally polarized Mott insulator transition; for largest $U/t$,
there is a metal to partially orbitally polarized Mott insulator transition.

For $J/U>0$ (see Figure \ref{fig:phase_diagram}, panels $b$-$f$),
the metal to insulator transition value of $U/t$ is no longer a monotonic
function of $n_{1\sigma}$. There is a first-order insulator to metal
transition in $\Delta$ around $n_{1\sigma}=1/2$, and the range increases
with $J/U$. For sufficiently large $J/U$ in the large polarization
region, the zero quasiparticle weight boundary coincides with the
first-order phase boundary. Additionally, there is an approximately
vertical boundary between the partially and largely orbitally polarized
Mott insulators, which becomes first-order for sufficiently large
$J/U$. Finally, the quasiparticle weight decreases as $n_{1\sigma}\rightarrow0$
for all $U/t>0$, and this is most easily seen for the larger values
of $J/U$. This behavior is expected given our scaling analysis in
Section \ref{subsec:Hamiltonian-for-the}, which suggests that for
$n_{1\sigma}<\alpha J^{2}$ the quasiparticle weight is zero, where
$\alpha$ is a positive constant. Given the numerical difficulty for
treating small values of $n_{1\sigma}$ (i.e. $n_{1\sigma}<0.01$),
it would be preferable to explore this regime treating $n_{1\sigma}$
as a small parameter, which would allow for an analytic evaluation
using $\discn=3$. Such an exercise would clearly answer whether or
not the Mott insulator exists for infinitesimal $U/t$ with fixed
$J/U>0$ in the large polarization limit. Nonetheless, the presence
of strong electronic correlations in the largely polarized regime
is clear. Therefore, crystals bearing $d$-electrons or $f$-electrons
which are nominally a band insulator, according to experiment or density
functional theory, may in reality be in this largely polarized regime
which has nontrivial electronic correlations. In future work, we will
investigate the doping dependence of this regime. 

\section{Conclusions and Future Outlook\label{sec:Conclusion-and-Future}}

In this work, we applied the recently developed VDAT within the SCDA
to the two-orbital Hubbard model in $d=\infty$. The SCDA is a self-consistent
algorithm to compute the total energy under the SPD using an iterative
approach, and this poses a serious technical challenge of how to efficiently
compute the derivatives of the total energy with respect to the variational
parameters. We surmounted this challenge using an iterative decoupled
minimization algorithm. At each iteration, the variational parameters
are updated using a local effective model and a collection of independent
effective models for the $k$-points, and the SCDA self-consistency
is maintained using a fixed point method. In addition to this minimization
algorithm, two formal developments were made to VDAT. First, we provided
a diagrammatic proof of the equivalence of integer time correlation
functions under the SPD to corresponding observables measured in the
compound space. Second, we identified the gauge symmetry of the SPD,
and we proposed various schemes for fixing the gauge freedom, which
is of practical importance for stabilizing the minimization within
the SCDA. 

Using the aforementioned formal and technical developments, we studied
the half filled two orbital Hubbard model in $d=\infty$ over a broad
range of parameter space in $U/t$, $J/U$, and $\Delta/t$ at zero
temperature. The computational cost of VDAT for this model is negligible,
requiring approximately one second on a single processor core at $\discn=3$
to solve the model for a given $U/t$, $J/U$, and $\Delta/t$. At
$\Delta=0$, we evaluated $\discn=2,3,4$, where $\discn=2$ recovers
the Gutzwiller approximation, and the results for $\discn=3,4$ only
exhibited very small differences, suggesting the results are largely
converged with respect to $\discn$. Given that increasing $\discn$
monotonically approaches the exact solution, $\discn=3,4$ should
be close to the exact solution, and therefore $\discn=3$ should serve
as a standard theory of Mott and Hund physics in the $d=\infty$ Hubbard
model. VDAT for $\discn>2$ confirms the previous Gutzwiller results
(i.e. $\discn=2)$ that the $U$ driven MIT for $\Delta=0$ (i.e.
$n_{1\sigma}=1/2$) and $J>0$ is first-order, and is continuous for
$J=0$. For $\Delta>0$ and $J>0$, VDAT for $\discn=3$ confirms
previous finite temperature DMFT results of a zero orbital susceptibility
region for $\Delta\apprle\sqrt{2}J$, and confirms previous conjectures
that the transition to finite susceptibility is sharp at zero temperature\cite{Werner2007126405}.
At intermediate values of $U/t$ and sufficiently large values of
$J/U$, there exists a first-order $\Delta$ driven MIT going from
a zero orbitally polarized Mott insulator to a partially polarized
metal, followed by a first-order transition to either a partially
or largely orbitally polarized Mott insulator. For large $U/t$, there
is a $\Delta$ driven transition from a partially to a largely orbitally
polarized Mott insulator, and this transition appears to be continuous
at small $J/U$ and first-order at large $J/U$. Finally, for nonzero
$J/U$, the quasiparticle weight decreases as $n_{1\sigma}\rightarrow0$
for all nonzero $U/t$, and this VDAT result is consistent with scaling
arguments in the large polarization limit. Detailed phase diagrams
of the quasiparticle weight as a function of $n_{1\sigma}$ and $U/t$
are presented at an unprecedented resolution. In summary, VDAT uncovered
qualitative physics which had not yet been resolved, and this is due
to the fact that VDAT operates at zero temperature and exactly evaluates
the SPD ansatz for arbitrary $U$, $J$, and $\Delta$. 

Analogous to DMFT, VDAT within the SCDA can be straightforwardly applied
in finite dimensions as a local approximation. Therefore, VDAT within
the SCDA at $\discn=3$ will likely become a \textit{de facto} standard
for probing the local physics of multiband Hubbard models at zero
temperature, delivering the quality of DMFT at a cost not far beyond
the Gutzwiller approximation. An obvious next step will be to combine
VDAT within the SCDA with DFT, in the same spirit of DFT plus Gutzwiller\cite{Deng2009075114}.
DFT+VDAT($\discn=3$) will have similar quality to DFT+DMFT at a cost
similar to DFT+Gutzwiller, and DFT+VDAT will have distinct advantages
over DFT+DMFT in that it naturally accesses zero temperature. 

Another important future direction will be executing VDAT within finite
dimensions. There are various approaches to extend DMFT to finite
dimensions, such as cluster DMFT\cite{Maier20051027,Kotliar2006865},
dual Fermions\cite{Rohringer2018025003}, the dynamical vertex approximation\cite{Rohringer2018025003},
etc, and it is clear that integer time analogues can be pursued within
VDAT. Given the massive speedup of VDAT within the SCDA relative to
DMFT, it seems likely that there will be a similar speedup when applying
VDAT to finite dimensions. Therefore, it seems possible that the SPD
at $\discn=3$ and beyond can be precisely evaluated using VDAT in
finite dimensions, allowing for a zero temperature solution that would
compete with all existing state-of-the-art methods for the single
band Hubbard model\cite{Leblanc2015041041}.

\section{Acknowledgments}

This work was supported by the Columbia Center for Computational Electrochemistry. 

\section{Appendix \label{sec:Appendix}}

Here we review some key properties of non-interacting many-body density
matrices and derive an explicit expression for $\barhat[Q]$. Consider
a generalized non-interacting many-body density matrix $\hat{\rho}_{0}=\exp\left(\boldsymbol{v}\cdot\hat{\boldsymbol{n}}\right)$,
where $\hat{\rho}_{0}$ is generalized in the sense that $\boldsymbol{v}$
is an arbitrary matrix. The following set of identities are useful
when evaluating the non-interacting integer time Green's function
\begin{align}
 & \hat{\rho}_{0}^{-1}\hat{a}_{i}^{\dagger}\hat{\rho}_{0}=\sum_{i'}\left[\bm{S}_{0}\right]_{ii'}\hat{a}_{i'}^{\dagger},\label{eq:appsim}\\
 & \hat{\rho}_{0}\hat{a}_{i}^{\dagger}\hat{\rho}_{0}^{-1}=\sum_{i'}\left[\bm{S}_{0}^{-1}\right]_{ii'}\hat{a}_{i'}^{\dagger},\label{eq:appsim2}\\
 & \hat{\rho}_{0}^{-1}\hat{a}_{i}\hat{\rho}_{0}=\sum_{i'}\hat{a}_{i'}\left[\bm{S}_{0}^{-1}\right]_{i'i},\label{eq:appsim3}\\
 & \hat{\rho}_{0}\hat{a}_{i}\hat{\rho}_{0}^{-1}=\sum_{i'}\hat{a}_{i'}\left[\bm{S}_{0}\right]_{i'i},\label{eq:appsim4}
\end{align}
where $\boldsymbol{S}_{0}=\exp\left(-\boldsymbol{v}^{T}\right)$.
In order to prove Eq. \ref{eq:appsim}, we first prove that it holds
for an infinitesimal $\tilde{\boldsymbol{v}}$ by directly evaluating
\begin{align}
 & \exp\left(-\tilde{\boldsymbol{v}}\cdot\hat{\boldsymbol{n}}\right)\hat{a}_{i}^{\dagger}\exp\left(\tilde{\boldsymbol{v}}\cdot\hat{\boldsymbol{n}}\right)=\hat{a}_{i}^{\dagger}+\left[\hat{a}_{i}^{\dagger},\tilde{\boldsymbol{v}}\cdot\hat{\boldsymbol{n}}\right]\nonumber \\
 & =\hat{a}_{i}^{\dagger}-\sum_{i'}[\tilde{\boldsymbol{v}}]_{i'i}\hat{a}_{i'}^{\dagger}=\sum_{i'}[\tilde{\boldsymbol{S}}_{0}]_{ii'}\hat{a}_{i'}^{\dagger},\label{eq:appinf}
\end{align}
where $\tilde{\boldsymbol{S}}_{0}=\exp\left(-\tilde{\boldsymbol{v}}^{T}\right)$.
For a finite $\boldsymbol{v}$, consider $\tilde{\boldsymbol{v}}=\boldsymbol{v}/N$
and iteratively apply Eq. \ref{eq:appinf} $N$ times with $N\rightarrow\infty$,
which proves Eq. \ref{eq:appsim}. Equations \ref{eq:appsim2}-\ref{eq:appsim4}
can then be derived from Eq. \ref{eq:appsim}. Using the preceding
identities, we can derive
\begin{equation}
\frac{\textrm{Tr}(\hat{\rho}_{0}\hat{a}_{i}^{\dagger}\hat{a}_{j})}{\textrm{Tr}(\hat{\rho}_{0})}=\left[\frac{\boldsymbol{1}}{\boldsymbol{1}+\boldsymbol{S}_{0}}\right]_{ij}.
\end{equation}

We now proceed to derive an explicit expression for $\barhat[Q]$,
and to simplify notation we consider $L=1$, though the derivation
is general. We begin by explicitly evaluating $\boldsymbol{g}_{Q}$
using Eq. 93 in Ref. \cite{Cheng2021195138}, resulting in
\begin{equation}
[\boldsymbol{g}_{Q}]_{\tau\tau'}=\frac{1}{2}\textrm{sign}(\tau'-\tau+\frac{1}{2}).
\end{equation}
Using Eqns. 112 and 113 from Ref. \cite{Cheng2021195138}, we obtain
\begin{equation}
\barhat[Q]=\exp\left(-\ln\left(\bm{S}_{Q}^{T}\right)\cdot\barhat[\boldsymbol{n}]\right)=\exp\left(\ln\left(\bm{S}_{Q}\right)\cdot\barhat[\boldsymbol{n}]\right),
\end{equation}
where $\boldsymbol{S}_{Q}=\boldsymbol{g}_{Q}^{-1}-\boldsymbol{1}$,
and the matrix elements are
\begin{equation}
[\boldsymbol{S}_{Q}]_{\tau\tau'}=-\delta_{\tau+1,\tau'}+\delta_{\tau-\discn+1,\tau'},\label{eq:appSq}
\end{equation}
which implies $\boldsymbol{S}_{Q}^{-1}=\boldsymbol{S}_{Q}^{T}=\boldsymbol{S}_{Q}^{\dagger}$
and $\barhat[Q]^{\dagger}=\barhat[Q]^{-1}$. We can now compute
\begin{equation}
\ln\bm{S}_{Q}=\sum_{\omega}\ln\left(-\lambda_{\omega}\right)v_{\omega}v_{\omega}^{T},
\end{equation}
where $\omega=\frac{2\pi}{\discn}m$ and $m=1,2,\dots,\discn$ and
$\lambda_{\omega}=\exp\left(i\left(\omega-\frac{\pi}{\discn}\right)\right)$
and
\begin{equation}
v_{\omega}=\frac{1}{\sqrt{\discn}}\left(1,\lambda_{\omega}^{1},\dots,\lambda_{\omega}^{\discn-1}\right)^{T}.
\end{equation}
The matrix elements of $\ln\boldsymbol{S}_{Q}$ can then be evaluated
as
\begin{align}
[\ln\bm{S}_{Q}]_{\tau\tau'} & =\frac{i}{\discn}\sum_{\omega}\left(\omega-\frac{\pi}{\discn}-\pi\right)\lambda_{\omega}^{\tau-\tau'}\\
 & =\begin{cases}
0 & \tau=\tau'\\
\frac{\pi}{\discn}\frac{1}{\sin\left(\pi\left(\tau-\tau'\right)/\discn\right)} & \tau\ne\tau'
\end{cases}.
\end{align}
Finally, using Eqns. \ref{eq:appsim}-\ref{eq:appsim4} and Eq. \ref{eq:appSq},
we obtain
\begin{align}
 & \barhat[Q]^{-1}\barhat[a]^{\dagger\left(\tau\right)}\barhat[Q]=-\barhat[a]^{\dagger\left(\tau+1\right)},\\
 & \barhat[Q]\barhat[a]^{\dagger\left(\tau\right)}\barhat[Q]^{-1}=-\barhat[a]^{\dagger\left(\tau-1\right)},\\
 & \barhat[Q]^{-1}\barhat[a]^{\left(\tau\right)}\barhat[Q]=-\barhat[a]^{\left(\tau+1\right)},\\
 & \barhat[Q]\barhat[a]^{\left(\tau\right)}\barhat[Q]^{-1}=-\barhat[a]^{\left(\tau-1\right)},
\end{align}
where we define $\barhat[a]_{\ell}^{\dagger(\discn+1)}\equiv-\barhat[a]_{\ell}^{\dagger(1)}$
and $\barhat[a]_{\ell}^{\dagger(0)}\equiv-\barhat[a]_{\ell}^{\dagger(\discn)}$.

%

\end{document}